\documentclass[reprint,nofootinbib,amsmath,amssymb,prd]{revtex4-2}
\usepackage{graphicx}
\usepackage{dcolumn}
\usepackage{bm}
\usepackage{color} 
\usepackage{amsmath,amssymb}
\usepackage{epsf}
\usepackage{color}
\usepackage[italicdiff]{physics}
\usepackage{mathtools}
\usepackage{enumitem}
\usepackage{tensor}
\usepackage{wrapfig}
\usepackage{footnote}
\usepackage{here, comment}
\usepackage{bbm}
\usepackage{ulem}

\usepackage[export]{adjustbox}

\usepackage[T1]{fontenc}

\begin{document}

\preprint{KEK-TH-2409}

\title{Entanglement distillation toward minimal bond cut surface in tensor networks}

\author{Takato Mori}
\email{moritaka@post.kek.jp}
\affiliation{
 KEK Theory Center, Institute of Particle and Nuclear Studies, High Energy Accelerator Research Organization (KEK), \\1-1 Oho, Tsukuba, Ibaraki 305-0801, Japan
}%
\affiliation{
 Department of Particle and Nuclear Physics, School of High Energy Accelerator Science, The Graduate University for Advanced Studies (SOKENDAI), \\1-1 Oho, Tsukuba, Ibaraki 305-0801, Japan
}%

\author{Hidetaka Manabe}%
 \email{manabe@acs.i.kyoto-u.ac.jp}
 \affiliation{%
 Department of Advanced Mathematical Sciences, Graduate School of Informatics, \\ Kyoto University, Yoshida-honmachi, Sakyo-ku, Kyoto, Kyoto 606-8501, Japan
}%

\author{Hiroaki Matsueda}
 \email{hiroaki.matsueda.c8@tohoku.ac.jp}
 \affiliation{%
 Department of Applied Physics, Graduate School of Engineering, Tohoku University, Sendai 980-8579, Japan
}%
\affiliation{%
 Center for Spintronics Integrated Systems, Tohoku University, Sendai 980-8577, Japan
}%

\date{\today}

\begin{abstract}
{
{In} tensor networks, a geometric operation of pushing a bond cut surface toward a minimal surface corresponds to entanglement distillation. 
{Cutting} bonds defines a reduced transition matrix on the bond cut surface and the associated quantum state naturally emerges from it. We justify this picture quantitatively by evaluating the trace distance between the maximally entangled states and the states on bond cut surfaces in {the} multi-scale entanglement renormalization ansatz (MERA) and matrix product states in a canonical form.} 
{Our numerical result for} the random {MERA} {is in a reasonable agreement with our proposal}. The result sheds new light on a deeper understanding of the Ryu-Takayanagi formula for entanglement entropy in holography {and {the} emergence of geometry from the entanglement structure}.
\end{abstract}

\maketitle


\section{\label{sec:intro}Introduction}

It has been a long-standing problem in {spacetime} physics to resolve the mysterious relationship between information and gravity. 
{In accordance with this fundamental mystery}, {an} 
efficient approach is to examine complementarity between quantum entanglement and geodesic structures in the context of {the} holographic principle~\cite{Susskind:1994vu,Maldacena:1997re}. {Originally, the principle 
{is a relationship between a certain class of {a} conformally invariant theory 
known as {the} holographic conformal field theory (CFT) 
and
spacetime with a constant negative curvature.}} 
The complementarity was strongly motivated by the so-called Ryu-Takayanagi (RT) formula, where 
{entanglement entropy} in holographic quantum field theory is proportional to the area of the minimal (extremal) surface in its gravity dual called the RT surface~\cite{Ryu:2006bv}. 
The RT formula is essentially a holographic extension of the famous Bekenstein-Hawking formula for 
black hole entropy~\cite{Bekenstein:1973ur,Bardeen:1973gs,Hawking:1975vcx}. A very important feature of {a} black hole is the presence of radiation of Hawking pairs inside and outside the event horizon. {Then,} the theory is described by the Bogoliubov transformation in superconductivity to connect both sides of the event horizon. Thus, the RT surface should be characterized by the condensation of entangled pairs from elementary objects, which have critical information about the holographic spacetime. Therefore, the characterization by extraction of the entangled pairs at the surface is crucial for a comprehensive understanding of the RT formula including 
{previous extensive} research. 

{
{In} quantum information theory, entanglement entropy can be defined operationally. {Entanglement} entropy of a state asymptotically equals 
the number of extractable Einstein–Podolsky–Rosen (EPR) pairs via local operations and classical communication 
in the limit of {the} large number of 
state {copies}. This procedure of 
extraction is called entanglement distillation.
To further clarify the information theoretic aspect of holography, it is important to understand how the RT formula is derived from the operation-based definition of entanglement entropy. However, the previous derivation of the RT formula~\cite{Lewkowycz:2013nqa} 
relies on the 
state-based definition and the relation to the operational definition remains unclear (although see \cite{Bao:2018pvs} for some progress regarding one-shot entanglement distillation). {The} bit thread {formalism}, a mathematically equivalent formulation of the RT formula, suggests EPR pairs across the RT surface~\cite{Freedman:2016zud}. While this picture supports the operational definition in holography, the physical origin of the EPR pairs is still unknown in contrast to the case of a black hole.}
In this paper, we address this issue in terms of quantum operational techniques {based on tensor networks, which we will describe 
{in the following}}.

The construction of {tensor networks,} variational wave functions that orient quantum information viewpoints, is deeply examined in statistical physics and condensed matter physics. 
{A tensor network} can be constructed by contracting internal bonds between tensors defined on each lattice site of our model system. The hidden {degrees of freedom} carried by the internal bonds represent how nonlocal quantum entanglement is shared between two distant sites. By controlling the dimension, we can sequentially increase the resolution of the variational optimization to obtain the true ground state. Mathematically, the network of the tensors is reformulated as projected entangled-pair states (PEPS)~\cite{Verstraete:2004cf}, in which we 
define maximally entangled states among artificial degrees of freedom on each bond and then {take} some physical mapping on each site. The well-known matrix product state (MPS) is a one-dimensional (1D) version of PEPS.

The PEPS construction 
{is closely related to} the 
{aforementioned} proposal for the condensation of entangled pairs at the RT surface in holography, except that the PEPS 
{do} not contain the extra holographic dimension. 
{
{However,} the multi-scale entanglement renormalization ansatz (MERA)~\cite{Vidal:2008zz} has an extra holographic dimension representing {the} coarse graining of information. Such tensor networks 
with an extra dimension have been proposed to be toy models of holography~\cite{Swingle:2009bg,Swingle:2012wq,Beny:2011vh,Pastawski:2015qua,Yang:2015uoa,Czech:2015kbp,SinaiKunkolienkar:2016lgg,Hayden:2016cfa,Evenbly:2017hyg,Jahn:2017tls,Bhattacharyya:2017aly,Bao:2018pvs,Qi:2018shh,Milsted:2018san,Steinberg:2020bef,Jahn:2020ukq,Jahn:2021kti} and have facilitated an information theoretic understanding of holography. To further gain insights, 
}
{
{coarse graining} is 
{key} to connect the boundary theory with the RT surface.} {In {the} MERA, the smaller edge of the exclusive causal cone corresponds to a discrete version of the RT surface.} 

A {recent} 
proposal for a better understanding of the RT surface 
{stated} that the maximally entangled states characterize the surface~\cite{Freedman:2016zud,Agon:2018lwq,Agon:2021tia,Rolph:2021hgz,Chen:2018ywy,Cui:2015pla}. The proposal suggests that the surface may emerge from entanglement distillation {by a deformation of the boundary}. 
{One of the goals in this paper is to provide a concrete {method} 
to 
{achieve} this 
{procedure} {in {the} MERA and discuss a possible extension to other tensor networks such as {MPS}}.} 

Motivated by the possible relationship with distillation and the minimal bond cut surface, we examine {geometric operations} in tensor networks with and without a holographic direction. 
{In 
special circumstances, previous literature {has} established the relation between the discrete version of the RT formula and a (one-shot) entanglement distillation in tensor networks. These tensor networks are perfect~\cite{Pastawski:2015qua} or special tree tensor networks~\cite{Bao:2018pvs,Lin:2020yzf,Yu:2020zwk,Lin:2020ufd}. Using the isometric property of their composing tensors, we can show the state equals 
a collection of EPR pairs across the minimal surface via {the} so-called greedy algorithm.}
However, these tensor networks are still inadequate 
{to achieve} 
{conformally invariant states, which are usually assumed in holography.} 
For instance, a correlation function in perfect tensor networks does not decay as the distance increases and its entanglement spectrum is flat. This is contradictory to the result for CFTs. 
Thus, we focus on MERA in this paper as it is known to efficiently approximate critical ground states. Furthermore, it has a capacity to express various wave functions via {a} variational optimization, which is also missing in the 
{holographic tensor network toy models} 
in 
previous literature. Despite {MERA} {being} neither {a} perfect nor {a} tree {tensor network}, our method 
{enables us to} discuss entanglement distillation in the {MERA}. Moreover, 
{we} claim that the methodology is also applicable to an MPS. 
There is no direct bulk/boundary correspondence in {MPS} since it {lives} on the lattice of our target model. However, when we define a partial system, a minimal bond cut surface can always be defined as the edge of the partial system. By appropriately distilling over each matrix, we can find a state close to the EPR pair. Our goal is to show analytical and numerical evidence for these procedures {
{in relation to} a minimal bond cut surface and EPR pairs}.

{In Sec. II, we describe our proposal of entanglement distillation 
{achieved} by a geometric procedure in {MERA} and quantify this using a trace distance. In Sec. III, we numerically demonstrate the procedure in the so-called random MERA. In Sec. IV, we extend our proposal to {MPS}. Finally, we summarize our work and discuss possible future directions.}

\section{\label{sec:HED}
Entanglement distillation for {MERA}}

\begin{figure}[h]
    \centering
    \includegraphics[width=0.9\linewidth]{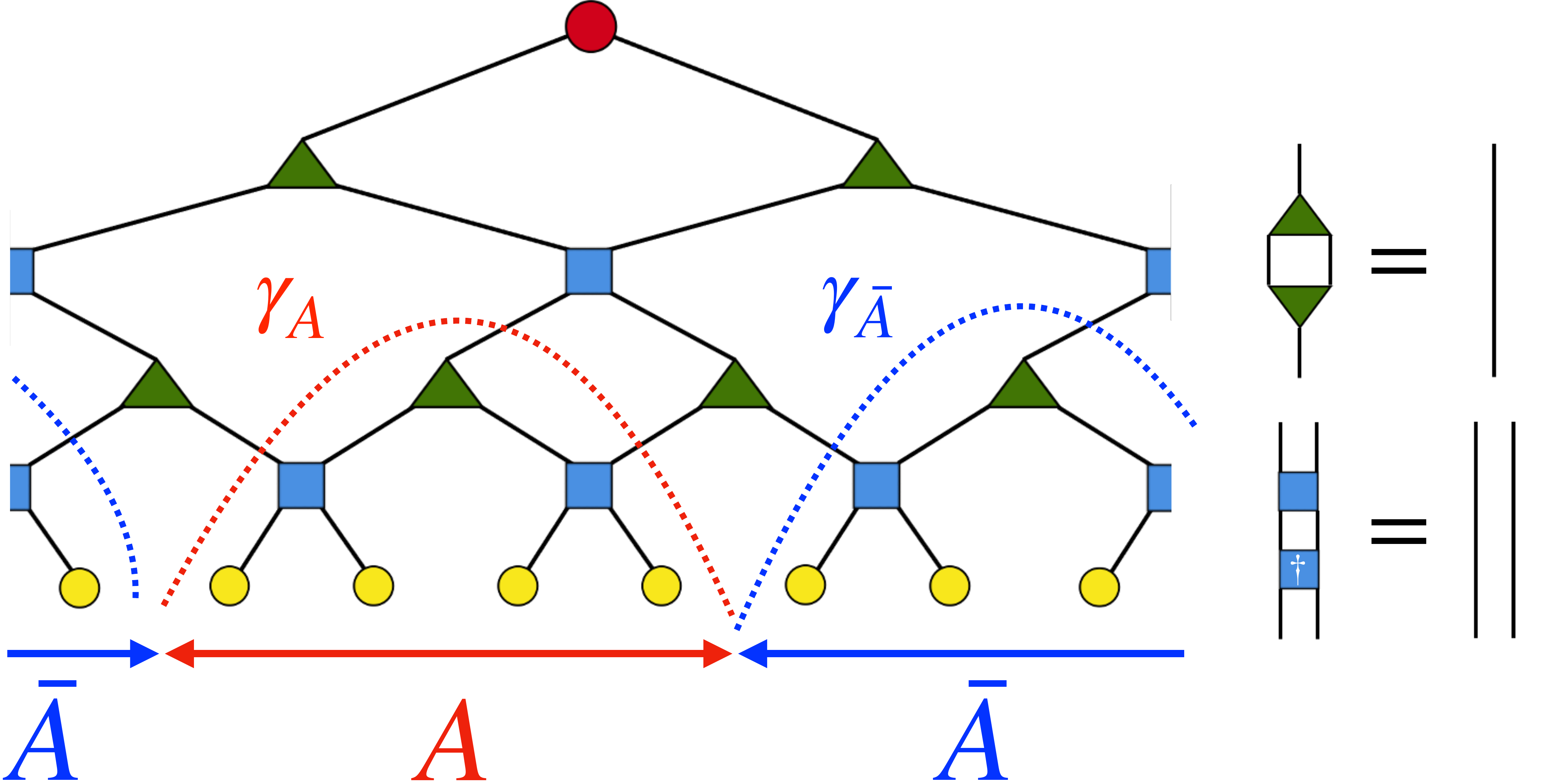}
    \caption{{A 
    MERA tensor network is composed of binary unitaries (blue squares), isometries (green triangles), and a top tensor (red circle). Yellow circles represent 
    physical indices. $A$ and $\bar{A}$ denote a subregion and its complement{, respectively}. For this symmetric bipartition, both $\gamma_A$ and $\gamma_{\bar{A}}$ become minimal bond cut surfaces $\gamma_\ast$.}}
    \label{fig:mera}
\end{figure}

\begin{figure}[h]
    \centering
    \includegraphics[width=\linewidth]{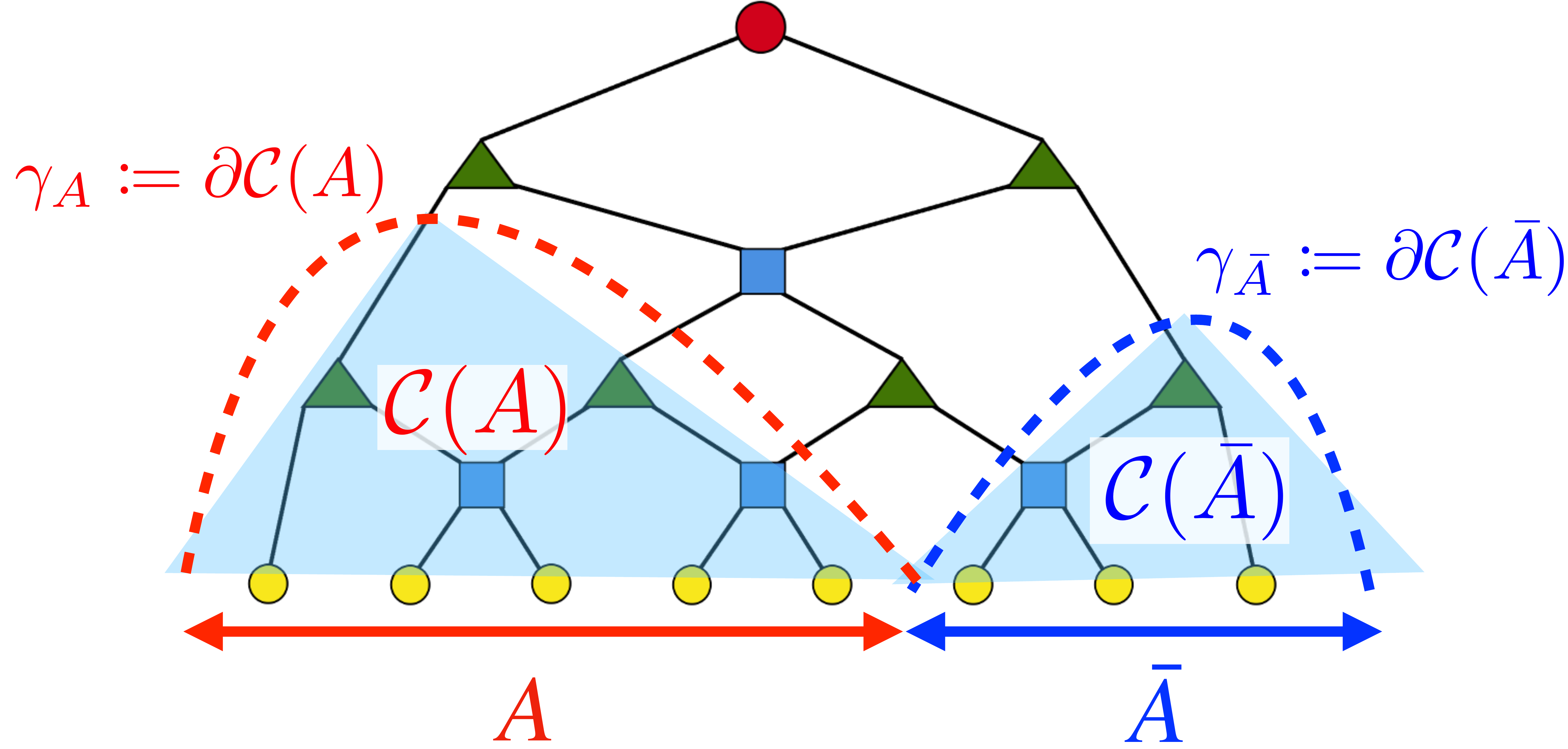}
    \caption{{{In a MERA tensor network}, the future or exclusive causal cone $\mathcal{C}(A)$ [$\mathcal{C}(\bar{A})$] of a subregion $A$ [$\bar{A}$] covers tensors that can 
    {affect} only $A$ [$\bar{A}$] seen from the top to 
    {bottom}. The edge of $\mathcal{C}(A)$ [$\mathcal{C}(\bar{A})$] is called a causal cut~\cite{Czech:2015kbp} or a minimal curve~\cite{Swingle:2009bg,Swingle:2012wq} and is denoted by $\partial\mathcal{C}(A)$ [$\partial\mathcal{C}(\bar{A})$]. In the {aforementioned} example, 
    the minimal bond cut surface $\gamma_\ast$ is given by $\gamma_{\bar{A}}$.}}
    \label{fig:causal-cone}
\end{figure}

{In this section, we define geometric operations in {a} tensor network, {{MERA} in particular,} and relate it with entanglement distillation.} 
We consider a binary MERA state $\ket{\Psi}$ represented by Fig.\ref{fig:mera}. It is composed of unitaries (blue squares), isometries (green triangles), and a top tensor (red circle). 
{The isometric regions shaded blue in Fig.\ref{fig:causal-cone}}
are called future or exclusive causal cones~\cite{Vidal:2008zz,Evenbly:2007hxg,evenbly2013quantum,Czech:2015kbp}.
We denote them by {$\mathcal{C}(A)$ for a subregion $A$ and $\mathcal{C}(\bar{A})$ for the complement $\bar{A}$.}
{Their} edges 
are denoted by $\gamma_A\equiv\partial\mathcal{C}(A)$ and $\gamma_{\bar{A}}\equiv\partial\mathcal{C}(\bar{A})$. We call the smaller one, a minimal bond cut surface $\gamma_\ast=\min(\gamma_A,\gamma_{\bar{A}})$. 
This surface $\gamma_\ast$ in {MERA} corresponds to the RT surface, a minimal surface in a holographic spacetime. From the PEPS perspective, there are EPR pairs across the surface. 
{Since isometries do not affect entanglement, the EPR pairs carry all of the entanglement of the state if all the projection tensors are isometries.} 
This is true for {a} perfect tensor {network}, which consists of isometries. 
In contrast, 
{{MERA} has nontrivial projection degrees of freedom carried by each tensor.} 
{As a result,} this naive 
{view} of EPR pairs across the surface 
{becomes} subtle.

The quantum correlation between $A$ and $\bar{A}$ is captured by entanglement entropy $S(\rho_A)=-\tr_A \rho_A \log\rho_A$, which is defined as the von Neumann entropy of the reduced density matrix $\rho_A=\tr_{\bar{A}}\dyad{\Psi}$. When the state is described by a MERA, given the fixed bond dimension $\chi$, entanglement entropy satisfies the following inequality:
\begin{equation}
    S(\rho_A)\le (\text{\# of bond cuts by }\gamma_\ast)\times \log\chi.
    \label{eq:TN-EE}
\end{equation}
When \eqref{eq:TN-EE} is saturated, it is interpreted as a discrete version of the RT formula.

\begin{figure*}[t]
    \centering
    \includegraphics[width=\linewidth]{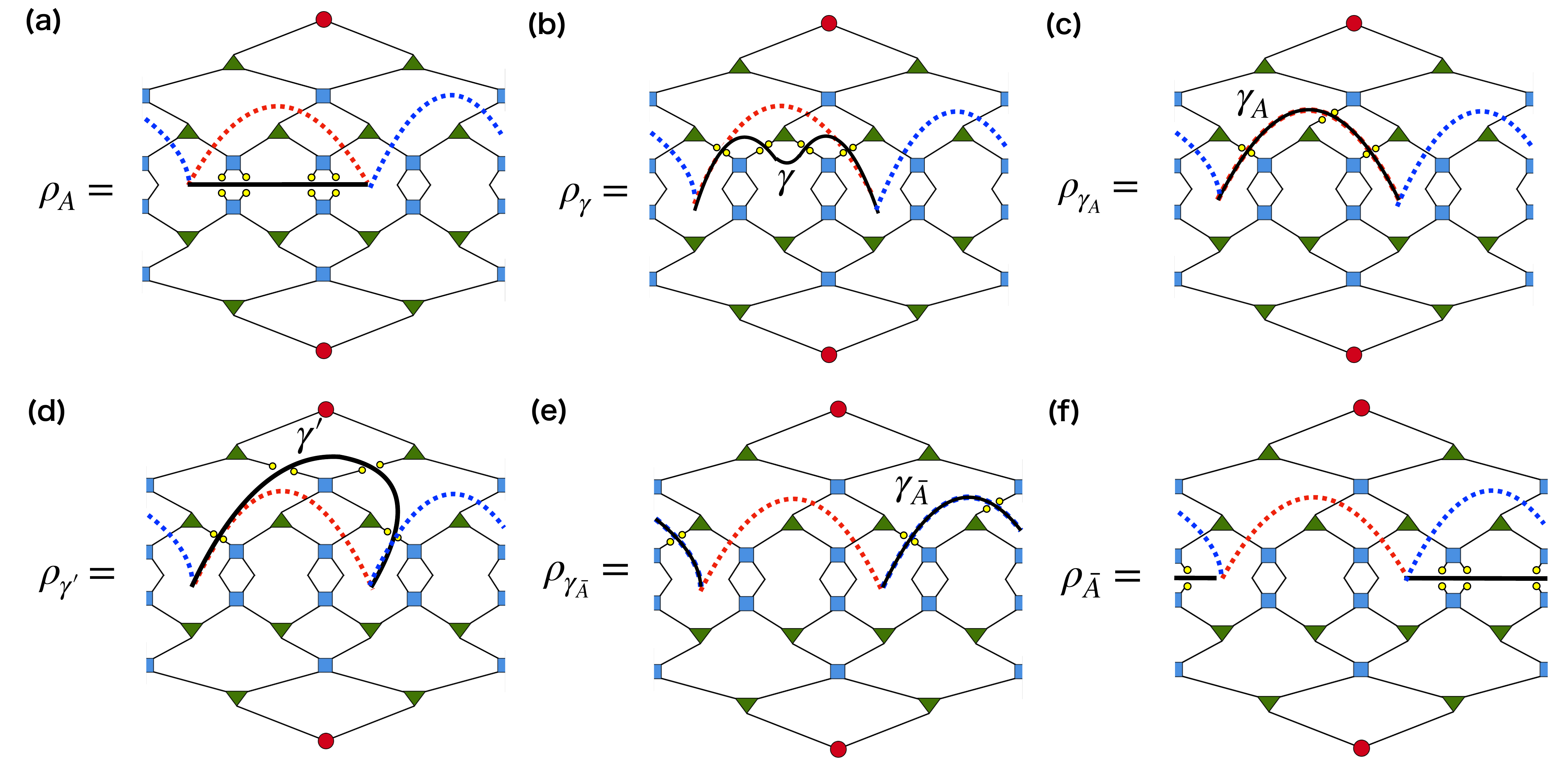}
    \caption{
    Reduced transition matrices corresponding to various foliations. 
    {When the subsystem $A$ is half of the whole system,} there are two minimal bond cut surfaces $\gamma_\ast=\gamma_A, \gamma_{\bar{A}}$.
    (a) Cutting the physical bonds of $A$ in $\braket{\Psi}$ gives $\rho_A$.
    (b) The foliation $\gamma$ is pushed toward $\gamma_A$.
    (c) The foliation equals 
    $\gamma_A$.
    (d) The foliation is pushed toward the other minimal bond cut surface $\gamma_{\bar{A}}$.
    (e) The foliation reaches $\gamma_{\bar{A}}$.
    (f) Finally, the foliation cuts the physical bonds in $\bar{A}$ and it gives the reduced density matrix $\rho_{\bar{A}}$.
    }
    \label{fig:HED}
\end{figure*}


In the following, we first present a 
way to define a state on 
{a surface across} internal bonds in {the} MERA. Then, such a state is shown to preserve the amount of entanglement 
{with an appropriate choice of} a family of bond cut surfaces. 
{As the minimal bond cut surface has the least number of bonds, we expect {the entanglement} per bond is concentrated to be {maximal}. Thus, we identify pushing a bond cut surface toward the minimal surface as entanglement distillation.} 
We quantify the process by examining the trace distance between each state and an EPR pair. 


    Given a MERA state $\ket{\Psi}$ (Fig.\ref{fig:mera}), its reduced density matrix $\rho_A$ for a subregion $A$ is obtained by cutting the physical bonds on $A$ in the norm $\braket{\Psi}$ as shown in Fig.\ref{fig:HED}~(a). 
    
    {In the following,} we consider a deformed surface $\gamma$ from $A$ such that the endpoints are common, $\partial\gamma=\partial A$. {This is a discrete version of the homology condition.} We call such a surface a \textit{foliation}. {As an initial condition}, we have $\gamma=A$. A minimal bond cut surface $\gamma_\ast$ {equals} 
    a foliation with a minimum number of bond cuts, i.e. 
    $\dim\mathcal{H}_\gamma\ge\dim\mathcal{H}_{\gamma_\ast}$, {where $\mathcal{H}_\gamma$ is the Hilbert space of bonds across $\gamma$}.
    
\begin{figure}[h]
    \centering
    \includegraphics[width=0.6\linewidth]{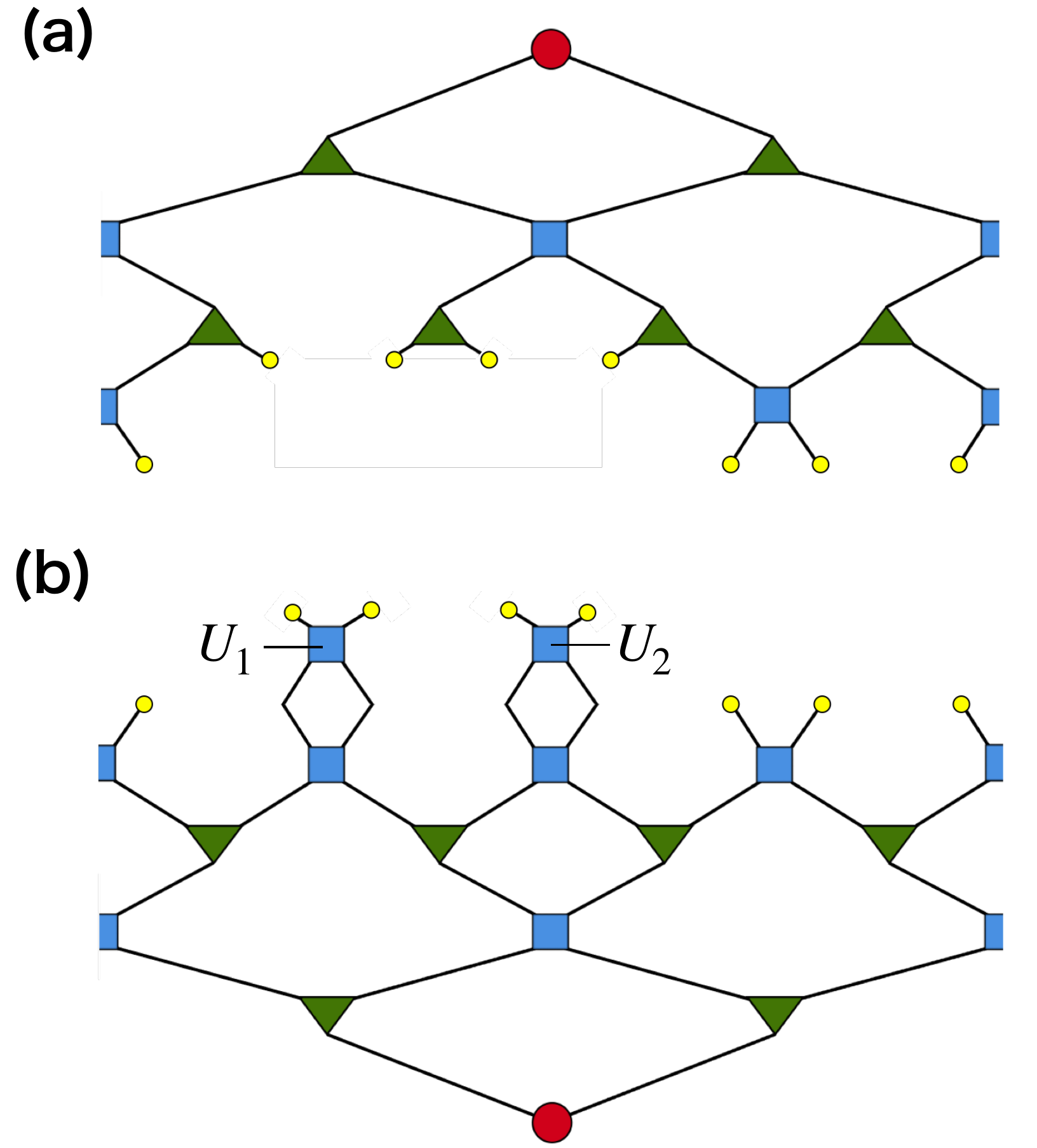}
    \caption{{When the foliation $\gamma$ is chosen as {shown in} Fig.\ref{fig:HED}~(b), $\ket{\Psi(\gamma)}$ is given by (a) and $\bra{\Phi(\gamma)}$ is given by (b). They are related to the original state $\ket{\Psi}$ by either removing or adding tensors $U_1\otimes U_2\in \mathcal{L}(\mathcal{H}_A)$.}}
    \label{fig:ketbra}
    \end{figure}

    Deforming $\gamma$ from $A$, we obtain a norm $\braket{\Psi}$ with bonds cut on $\gamma$. For example, when we choose a foliation $\gamma$ {as} shown in Fig.\ref{fig:HED} (b), the tensor network defines a reduced transition matrix~\cite{Nakata:2020luh}
    \begin{equation}
        \rho_\gamma = \tr_{\bar{A}}\left(\ket{\Psi(\gamma)} \bra{\Phi(\gamma)} \right)\in \mathcal{L}(\mathcal{H}_\gamma),
        \label{eq:red-trans}
    \end{equation}
    {where $\mathcal{L}(\mathcal{H})$ denotes a set of linear operators on a Hilbert space $\mathcal{H}$.}
    Fig.\ref{fig:ketbra} 
    {shows} the states $\ket{\Psi(\gamma)}\in \mathcal{H}_\gamma \otimes \mathcal{H}_{\bar{A}}$ and $\bra{\Phi(\gamma)}\in \mathcal{H}^\ast_\gamma \otimes \mathcal{H}^\ast_{\bar{A}}$. It immediately follows that $\braket{\Phi(\gamma)}{\Psi(\gamma)}=\tr \rho_\gamma=1$ for an arbitrary foliation $\gamma$. 
    $\bra{\Phi(\gamma)}$ and $\left|\Psi(\gamma)\right>$ are created by adding and removing tensors $M_\gamma$ bounded by $A$ and $\gamma$ in the tensor network representation:
    \begin{equation}
        \begin{aligned}
        \bra{\Phi(\gamma)}&=\bra{\Psi} M_\gamma\\
        M_\gamma \ket{\Psi(\gamma)} &= \ket{\Psi}.
        \end{aligned}
        \label{eq:braket-rel}
    \end{equation}
    For example, if we consider a configuration {shown in} Fig.\ref{fig:HED}~(b), $M_\gamma=U_1\otimes U_2$ where $U_{1,2}$ are {shown} 
    in Fig.\ref{fig:ketbra}~(b). 
    
    Using the relation \eqref{eq:braket-rel}, we can show {that} any reduced transition matrices $\rho_\gamma$ have common positive eigenvalues with the original reduced density matrix $\rho_A$. This can be shown as follows. We denote $\tr_{\bar{A}} \left(\ket{\Psi(\gamma)} \bra{\Psi} \right)$ by $S_\gamma$ and the positive eigenvalues and eigenvectors of $\rho_\gamma$ are denoted by $\{\lambda_n\}_n$ and $\{\ket{n}_\gamma\}_n$. Then,
    \begin{equation}
        \rho_\gamma \ket{n}_\gamma = S_\gamma M_\gamma \ket{n}_\gamma = \lambda_n \ket{n}_\gamma.
        \label{eq:gamma}
    \end{equation}
    By multiplying $M_\gamma$ from {the} left, we obtain
    \begin{equation}
        M_\gamma S_\gamma M_\gamma \ket{n}_\gamma = \lambda_n M_\gamma \ket{n}_\gamma
    \end{equation}
    whereas $M_\gamma S_\gamma = \tr_{\bar{A}} \left(M_\gamma\ket{\Psi(\gamma)} \bra{\Psi} \right)=\rho_A$ from \eqref{eq:braket-rel}. Since \eqref{eq:gamma} is by definition nonzero, $M_\gamma \ket{n}_\gamma \neq 0$. Therefore the positive eigenvalues of $\rho_A$ coincide with those of $\rho_\gamma$ for {an} arbitrary $\gamma$.
    

    {Since $\rho_A$ and $\rho_\gamma$ share common positive eigenvalues, it immediately follows that} 
    the von-Neumann entropy of a reduced transition matrix $S(\rho_\gamma)$ 
    {known} as pseudo entropy~\cite{Nakata:2020luh}, {equals} 
    entanglement entropy:
    \begin{equation}
    S(\rho_\gamma)=S(\rho_A),\quad \forall \gamma \quad \mathrm{s.t.}\quad \partial\gamma=\partial A.
    \label{eq:pseudo}
    \end{equation}
    This identity is interpreted in two ways. 
    {The first} is a 
    {type} of {the} bulk/boundary correspondence like the RT formula. While the right-hand side represents the {entanglement} entropy of the boundary quantum state $\ket{\Psi}$, the left-hand side is given as a function of an operator in the bulk. 
    {The second} is interpreted as a conservation of entanglement during the deformation of $\gamma$. From the PEPS perspective, $S(\rho_\gamma)$ effectively counts the amount of entanglement carried by bonds across $\gamma$. Then, the equality \eqref{eq:pseudo} indicates the amount of entanglement across each foliation is retained during the deformation of $\gamma$. 
    
    Throughout the procedure, the number of 
    {bond} cuts at $\gamma$ changes and it minimizes at a minimal bond cut surface $\gamma_\ast$. Thus, the diluted entanglement over $\ket{\Psi}$ is concentrated into a smaller number of strongly entangled bonds across $\gamma_\ast$. Next, we evaluate the degree of this concentration in terms of the trace distance.
    


{Before moving on, let us comment on} the 
{similarities and differences} between our procedure and previous proposals. In our procedure, we define 
{a reduced transition matrix} on each foliation and identify pushing the foliation as entanglement distillation. 
{Compared with} the previous studies of entanglement distillation in holography~\cite{Pastawski:2015qua,Bao:2018pvs}, 
{pushing the foliation} can be regarded as a 
{type} of operator pushing. In~\cite{Pastawski:2015qua}, an operator pushing of 
{an} operator $O$ through an isometry $V_{\mathrm{iso}}$ is defined by
\begin{equation}
    OV_{\mathrm{iso}}=V_{\mathrm{iso}}\Tilde{O},
\end{equation}
where $\Tilde{O}=V^\dagger_{\mathrm{iso}} O V_{\mathrm{iso}}$. 
{While $O$ is usually state-independent,} 
in our procedure, the pushed operator is 
{the reduced transition matrix defined from the state. The mapping between two reduced transition matrices $\rho_\gamma$ and $\rho_{\gamma^\prime}$ on the foliations $\gamma$ and $\gamma^\prime$ respectively is an operator pushing, i.e.}
\begin{equation}
    \rho_\gamma M = M \rho_{\gamma^\prime},
\end{equation}
where 
{$M$ represents tensors bounded by $\gamma$ and $\gamma^\prime$.} 
{Although our procedure can be interpreted as a 
{type} of operator pushing, one} 
important difference is that $M$ is not necessarily isometric while $V_{\mathrm{iso}}$ was assumed to be isometric or unitary. 
{
This difference arises because our procedure deals with a reduced transition matrix rather than a state vector. For a state vector, the only operations 
{that} preserve entanglement entropy are isometry and unitary {ones}. This requirement severely restricts possible tensor network states. They must be composed of perfect~\cite{Pastawski:2015qua} tensors or dual unitaries~\cite{Bertini:2018fbz,Bertini:2019lgy} or isometric tree tensor networks~\cite{Bao:2018pvs,Lin:2020yzf,Yu:2020zwk,Lin:2020ufd}. Such states can be distilled by removing the composing tensors 
{by applying} a greedy algorithm. 
In our procedure, we deal with a reduced transition matrix. The operations 
{that} preserve entanglement \eqref{eq:pseudo} are not limited to isometries. In this way, we can consider entanglement distillation using reduced transition matrices on various bond cut surfaces in a more general tensor network like {MERA}, which has nonisometric $M$.} 
{This} enables us to consider a state on an arbitrary bond cut surface even beyond the region a greedy algorithm can reach (called a bipartite residual region~\cite{Pastawski:2015qua} or causal shadow~\cite{Lewkowycz:2019xse} in the literature) while retaining the amount of entanglement $S(\rho_\gamma)$.

To evaluate how much entanglement is distilled from the original state $\ket{\Psi}$, we should quantify the closeness of a properly defined state across $\gamma$ to a maximally entangled state (the EPR pair).
However, since $\rho_\gamma$ is an operator, we cannot compare it with the EPR state 
{directly}. Thus, we define a distilled state on $\gamma$ as a state vector in $\mathcal{H}_\gamma\otimes\mathcal{H}_\gamma $ using the same idea with the purification,
\begin{equation}
    \ket{\rho_\gamma^{1/2}}\equiv {\mathcal{N}_\gamma} \sqrt{\dim\mathcal{H}_\gamma}
    (\rho_\gamma^{1/2}\otimes \mathbf{1})
    \ket{\mathrm{EPR}_\gamma},
    \label{eq:purif}
\end{equation}
where $\mathcal{N}_\gamma=\Big[{\tr(\rho_\gamma^{\dagger\, 1/2}\rho_\gamma^{1/2})}\Big]^{-1/2}$ and 
$\ket{\mathrm{EPR}_\gamma}=({\dim\mathcal{H}_\gamma})^{-1/2} \sum_{i=1}^{\dim\mathcal{H}_\gamma} \ket{i}\otimes\ket{i}$.
Then, we can define the closeness between the distilled {and EPR states} 
as the trace distance between them:
\begin{equation}
    D_\gamma\equiv \sqrt{1-\big|{\langle \mathrm{EPR}_\gamma | \rho_\gamma^{1/2}\rangle}\big|^2}.
    \label{eq:trace_distance}
\end{equation}
{On the basis of} this trace distance, we propose that the minimal bond cut surface $\gamma_\ast$ provides entanglement distillation such that $\ket{\rho_{\gamma}^{1/2}}$ becomes closest to the EPR pair $\ket{\mathrm{EPR}_\gamma}$ among other foliations $\gamma$.

For a later discussion, let us further rewrite \eqref{eq:trace_distance}. 
First, $\rho_A$ is represented by
\begin{equation}
    [\rho_A]_{IJ}=\sum_{\alpha^\prime=1}^r S_{I\alpha^\prime} \sigma_{\alpha^\prime}^2 S^\dagger_{\alpha^\prime J},
    \label{eq:svd}
\end{equation}
where $S$ is an isometry, $\sigma$ is a singular value matrix of $\ket{\Psi}$, and $r$ is the Schmidt rank. Note that $r\le \dim\mathcal{H}_{\gamma_\ast}$. 
{Then, as the positive eigenvalues are common between $\rho_\gamma$ and $\rho_A$,} 
the inner product in $D_\gamma$ can be written as
\begin{align}
    \langle \mathrm{EPR}_\gamma | \rho_\gamma^{1/2}\rangle 
    &=\frac{\mathcal{N}_\gamma}{\sqrt{\dim\mathcal{H}_{\gamma}}}
    \tr \rho_\gamma^{1/2} \nonumber\\
    &=\frac{\mathcal{N}_\gamma}{\sqrt{\dim\mathcal{H}_{\gamma}}}
    \sum_{\alpha^\prime=1}^r \sigma_{\alpha^\prime} \label{eq:trace-dist}\\
    &\le \frac{\mathcal{N}_\gamma}{\sqrt{\dim\mathcal{H}_{\gamma}}} r^{1/2} \sqrt{\sum_{\alpha^\prime=1}^r \sigma_{\alpha^\prime}^2} \nonumber\\
    &=\mathcal{N}_\gamma\sqrt{\frac{r}{\dim\mathcal{H}_{\gamma}}}.
    \label{eq:tr-dist-ineq}
\end{align}
The last line comes from the normalization $\tr\rho_A=1$. The inequality is saturated only when $\sigma\propto \mathbf{1}$.
We can further rewrite \eqref{eq:trace-dist} in terms of the $n$-th R\'enyi entropy
\[
S_n\equiv \frac{1}{1-n}\log\tr\rho_A^n =\frac{1}{1-n}\log \sum_{\alpha^\prime=1}^r \sigma_{\alpha^\prime}^{2n}.
\]
Since $S_{1/2}=2\log \sum_{\alpha^\prime=1}^r \sigma_{\alpha^\prime}$, \eqref{eq:trace-dist} is rewritten as
\begin{equation}
    \big|{\langle \mathrm{EPR}_\gamma | \rho_\gamma^{1/2}\rangle}\big|^2=\frac{\mathcal{N}_\gamma^2}{{\dim\mathcal{H}_{\gamma}}}e^{{S_{1/2}}}.
    \label{eq:trace-dist-renyi}
\end{equation}
In any cases, the $\gamma$-dependence in $D_\gamma$ only appears through $\mathcal{N}_\gamma$ and $\dim\mathcal{H}_\gamma$.

When $\gamma\subset \mathcal{C}(A)\cup \mathcal{C}(\bar{A})$, $M_\gamma$ is either 
isometric or unitary. {Thus, we can apply a standard greedy algorithm 
in this case. Since the tensors inside the causal cones are reduced to {an} identity after contractions,  a removal of tensors in $\mathcal{C}(A)\cup \mathcal{C}(\bar{A})$ from a state is equivalent to pushing the foliation in $\mathcal{C}(A)\cup \mathcal{C}(\bar{A})$. This means we can perform entanglement distillation {that is} perfectly consistent with the previous proposals. Let us see this from the view point of the trace distance.} By contracting isometries and unitaries in the MERA, this 
{indicates that}
    \begin{equation}
        \rho_\gamma^\dagger=\rho_\gamma
        \Rightarrow\mathcal{N}_\gamma=1.
        \label{eq:density-matrix}
    \end{equation}

From these expressions, the following statements can be derived for $\forall \gamma\subset \mathcal{C}(A)\cup\mathcal{C}(\bar{A})$. 
First, the inner product \eqref{eq:trace-dist} monotonically increases as we push $\gamma$ toward a minimal bond cut surface $\gamma_\ast$. This is because $\log\dim\mathcal{H}_\gamma$ is proportional to the number of bonds cut by $\gamma$. Then, from the definition \eqref{eq:trace_distance}, the trace distance monotonically decreases
\begin{equation}
    D_{\gamma^\prime} - D_\gamma < 0
    \label{eq:trace-dist-diff}
\end{equation}
as we push $\gamma$ to $\gamma^\prime$ toward a minimal bond cut surface $\gamma_\ast$. 
Second, 
\begin{equation}
    \gamma\neq\gamma_\ast \Rightarrow D_\gamma>0
    \label{eq:trace-pos}
\end{equation}
since from \eqref{eq:tr-dist-ineq}
\begin{equation}
    \langle \mathrm{EPR}_\gamma | \rho_\gamma^{1/2}\rangle \le \sqrt{\frac{r}{\dim\mathcal{H}_\gamma}}<1,
\end{equation}
where we used $r\le\dim\mathcal{H}_{\gamma_\ast}<\dim\mathcal{H}_\gamma$. The first statement \eqref{eq:trace-dist-diff} supports distilling a state closer to the EPR state by pushing $\gamma$ toward $\gamma_\ast$. The second statement \eqref{eq:trace-pos} indicates that we cannot have $D_\gamma=0$ (distillation of the EPR pair) unless $\gamma=\gamma_\ast$. The vanishing trace distance is equivalent to either a flat entanglement spectrum
\begin{equation}
    r=\dim\mathcal{H}_{\gamma_\ast}\qq{and}\sigma\propto\mathbf{1}
    \label{eq:schmidt}
\end{equation}
from \eqref{eq:tr-dist-ineq} or 
\begin{equation}
    S_{1/2}=\log\dim\mathcal{H}_{\gamma_\ast}
    \label{eq:renyi-1/2}
\end{equation}
from \eqref{eq:trace-dist-renyi}, which is expected from holography~\cite{Bao:2018pvs}.

\section{Numerical results for {random MERA}}
In this section, we demonstrate the 
{aforementioned} procedure of entanglement distillation in a particular MERA called the random MERA by again calculating a trace distance. The random MERA is prepared with Haar random unitaries {$U^{\alpha\beta}_{\gamma\delta}$}, where each index runs from $1$ to $\chi$. Isometries are given by $W^\alpha_{\beta\gamma}=U^{\alpha\, 1}_{\beta\gamma}$ and the top tensor is given by $T_{\alpha\beta}=U^{11}_{\alpha\beta}$.

The random MERA is particularly suitable to 
verify our {proposal of entanglement distillation}. Preceding studies have pointed out that a random tensor network can saturate \eqref{eq:TN-EE} in the large bond dimension limit, realizing a discrete version of the RT formula~\cite{Hayden:2016cfa,Swingle:2012wq,Qi:2018shh,Kudler-Flam:2019wtv}. 
{The goal of our method is to} extract pure EPR pairs on $\gamma_\ast$ in this limit as expected from holography, {and more importantly, whether this way of entanglement distillation really works even in a finite bond dimension, which is less trivial}.

Numerical calculations have been done for $8$-site and $16$-site random MERAs with bond dimension $\chi$. {For the $8$-site MERA,} we choose a subregion $A$ and foliations as shown in Fig{s}.\ref{fig:mera} and 
\ref{fig:HED}. 
{For the $16$-site MERA, we choose $6$-site and $8$-site subregions and foliations at their minimal bond cut surfaces.} 
Then, we calculate the trace distance \eqref{eq:trace_distance} to 
{investigate} the closeness between each state and the EPR state. We change the value of $\chi$ to see the 
{trend} of distillation in the large-$\chi$ limit. Tensor network contractions were performed using quimb~\cite{Gray2018} and cotengra~\cite{Gray2021}.

\begin{figure}[h]
    \centering
    \includegraphics[width=0.9\linewidth]{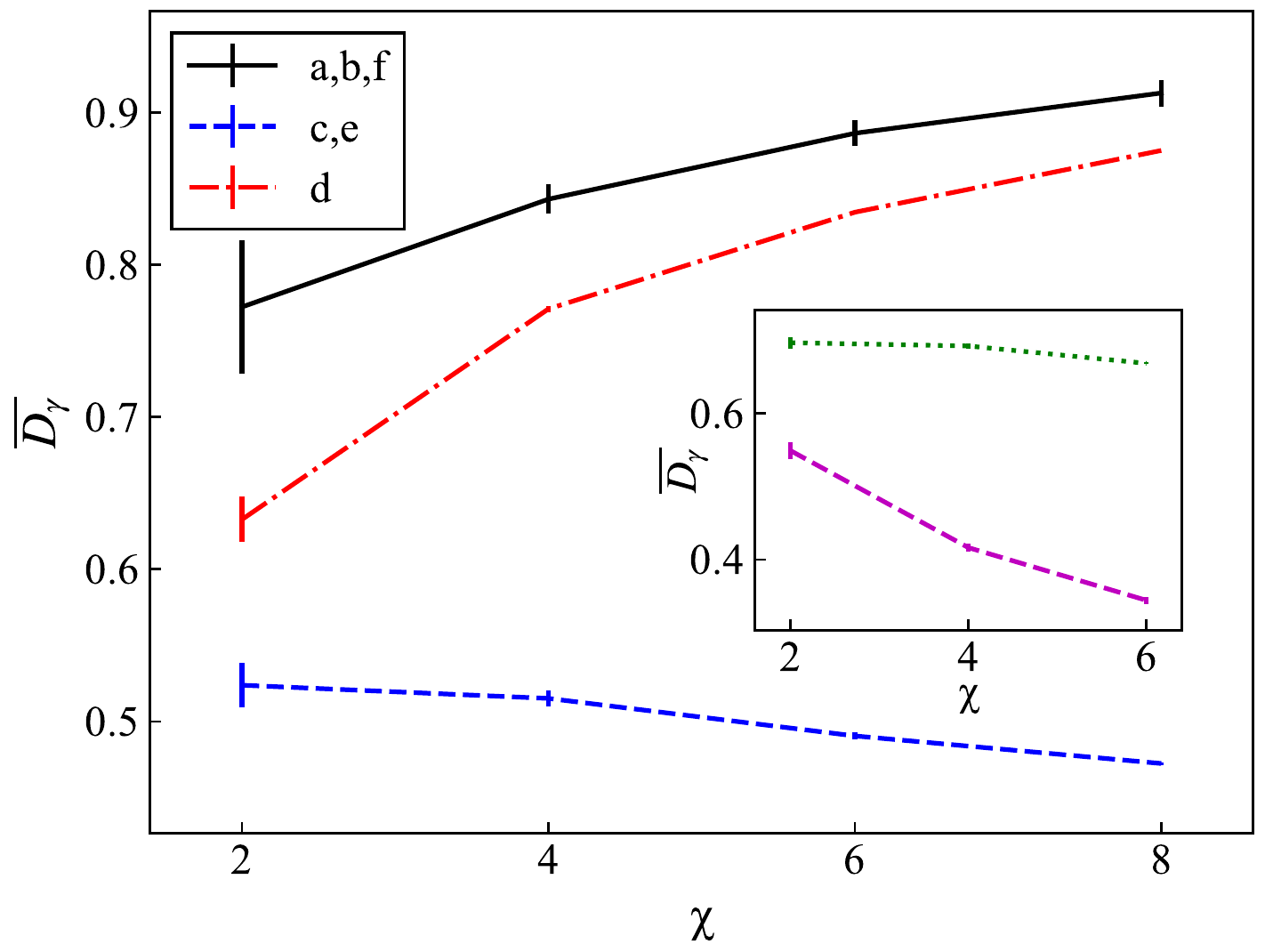}
    \caption{
    {Random}-averaged trace distance  $\overline{D_\gamma}$ 
    for each foliation in Fig.\ref{fig:HED} with respect to the bond dimension $\chi$ for the $8$-site random MERA. The inset is $\overline{D_\gamma}$ on the minimal bond cut surface for the $16$-site random MERA (green dotted line with an $8$-site subregion and purple dashed line with a $6$-site subregion).  
    }
    \label{fig:trace_distance}
\end{figure}

Fig.\ref{fig:trace_distance} shows the random-averaged trace distance $\overline{D_\gamma}$ for each foliation $\gamma$ in the $8$-site random MERA. Each $\overline{D_\gamma}$ is calculated using {ten} samples. The trace distances for the states on 
foliation{s} (a) and (b) are the same due to the equivalence up to a unitary transformation on $\rho_\gamma$. The distances for the states on (a) and (f) are also the same 
{as $A$ and $\bar{A}$ are complement to each other in the pure state.}
It is the same for (c) and (e) {which are related via a common unitary transformation from (a) and (f), respectively}. The state on 
{foliation} (d) corresponding to neither $\rho_{A}$, $\rho_{\bar{A}}$ nor $\rho_{\gamma_\ast}$ has a trace distance in between others. {Note that any greedy algorithms can never reach 
{foliation} (d) but our method enables us to compute the trace distance even for such a case in a well-defined manner.} 
We can see the foliation  $\gamma=\gamma_\ast$ (c,e) exhibits the smallest trace distance among all the foliations for bond dimensions from 2 to 8. 
The trace distances for (c,e) monotonically decrease as the bond dimension increases, which is consistent with~\cite{Hayden:2016cfa}. 
{These 
{trends} are} also seen in the situation of the $16$-site random MERA (Fig.\ref{fig:trace_distance} inset). This indicates this distillation procedure succeeds on the minimal bond cut surfaces $\gamma_\ast$ for each bond dimension {even when the bond dimension is not large}. 
{However,}
the trace distance on the other foliations increases 
as we increase the bond dimension. In this way, the minimal bond cut surface can be characterized from the perspective of distillation.

The behavior in the large-$\chi$ limit can be analytically understood as follows. The previous study~\cite{Hayden:2016cfa} found
\begin{equation}
    \lim_{\chi\rightarrow\infty}\overline{S_n}=\log\dim\mathcal{H}_{\gamma_\ast}
\end{equation}
for a non-negative integer. Assuming its analytical continuation to $n=1/2$ 
\begin{equation}
    \lim_{\chi\rightarrow\infty}\overline{S_{1/2}}=\log\dim\mathcal{H}_{\gamma_\ast},
\end{equation}
holds as expected from holography~\cite{Hayden:2016cfa}, \eqref{eq:trace-dist-renyi} and the Jensen's inequality leads
\begin{align}
    \lim_{\chi\rightarrow\infty}\overline{\big|{\langle \mathrm{EPR}_{\gamma_\ast} | \rho_{\gamma_\ast}^{1/2}\rangle}\big|^2} &=
    \lim_{\chi\rightarrow\infty}\frac{1}{\dim\mathcal{H}_{\gamma_\ast}}\overline{\exp(S_{1/2})} \nonumber\\
    &\ge\lim_{\chi\rightarrow\infty}\frac{1}{\dim\mathcal{H}_{\gamma_\ast}}\exp({\overline{S_{1/2}}})
    = 1.
\end{align}
Since the inner product between normalized states is at most one, we can conclude the distilled state approaches the EPR state for a large bond dimension. Even at a finite $\chi$, the existence of a gap between the distance for $\gamma=\gamma_\ast$ and others is consistent with \eqref{eq:trace-dist-diff}.

\section{Entanglement distillation for matrix product states}
Numerically we have seen our distillation method indeed works for the random MERA. To look for a possible extension to other classes of tensor networks{,} 
{we} focus on {MPS}, which belongs to a different criticality from {MERA}.

\begin{figure}[h]
    \centering
    \includegraphics[width=\linewidth]{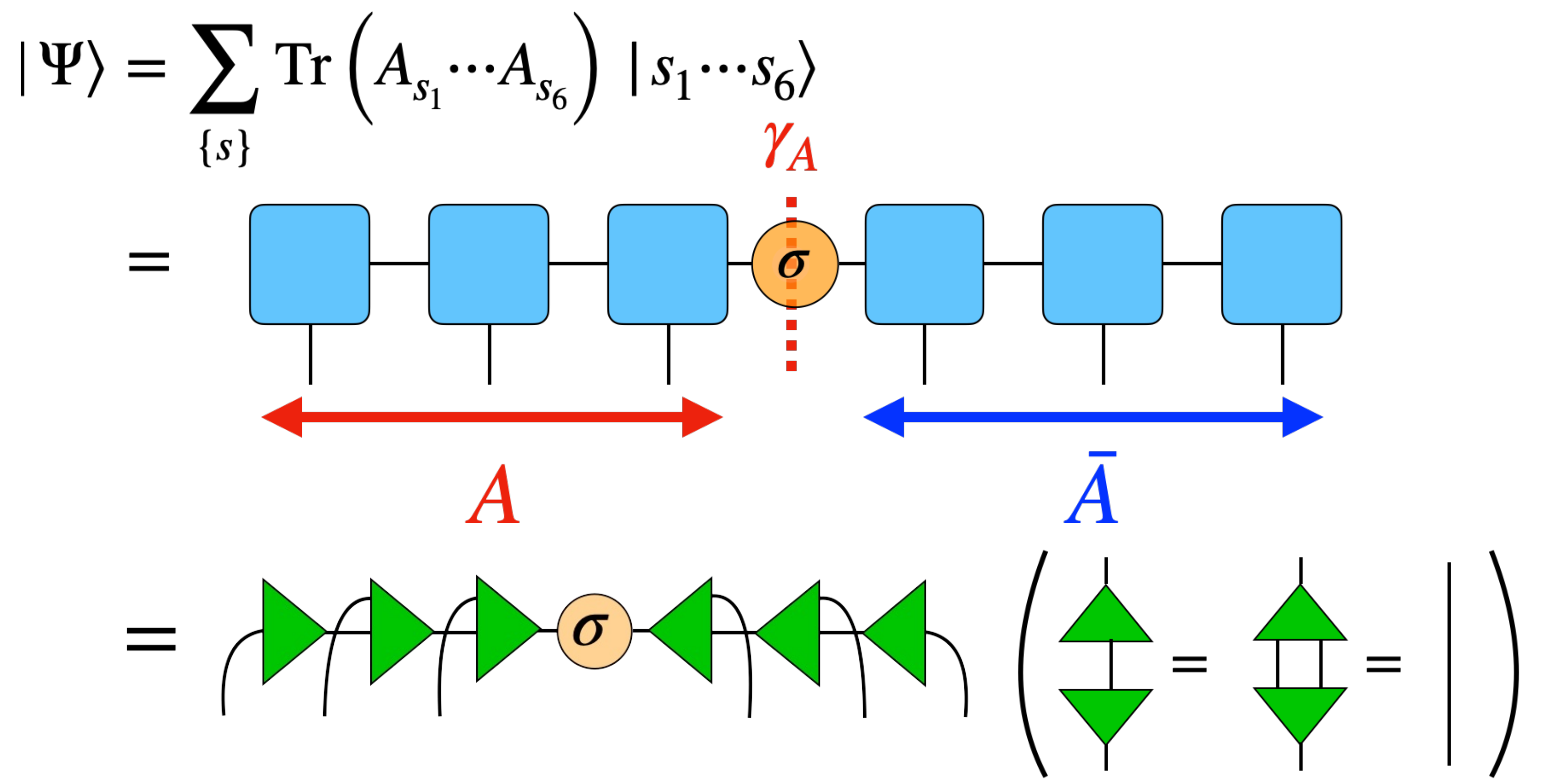}
    \caption{A matrix product state $\ket{\Psi}$ is shown in the first and second lines. The third line represents its mixed canonical form~\cite{SCHOLLWOCK201196} via a successive singular value decomposition for each matrix. Every matrix in the form is isometric (green triangle) and the singular value matrix $\sigma$ is placed in the center. The boundary between $A$ and $\bar{A}$ is denoted by $\gamma_A$.}
    \label{fig:MPS}
\end{figure}

{An} MPS with open boundaries is shown in {the first and second lines in} Fig.\ref{fig:MPS}. For simplicity, we focus on the case when the subregion $A$ is on the left and its complement $\bar{A}$ is on the right. The boundary between $A$ and $\bar{A}$ is denoted by $\gamma_A$. The bond dimension for internal bonds is denoted by $\chi$. We can always transform an MPS in a so-called mixed canonical form~\cite{SCHOLLWOCK201196}, shown in the third line of Fig.\ref{fig:MPS}, by a successive singular value decomposition of matrices. In the mixed canonical form, every matrix is isometric except for the singular value matrix $\sigma$.

\begin{figure}[h]
    \centering
    \includegraphics[width=\linewidth]{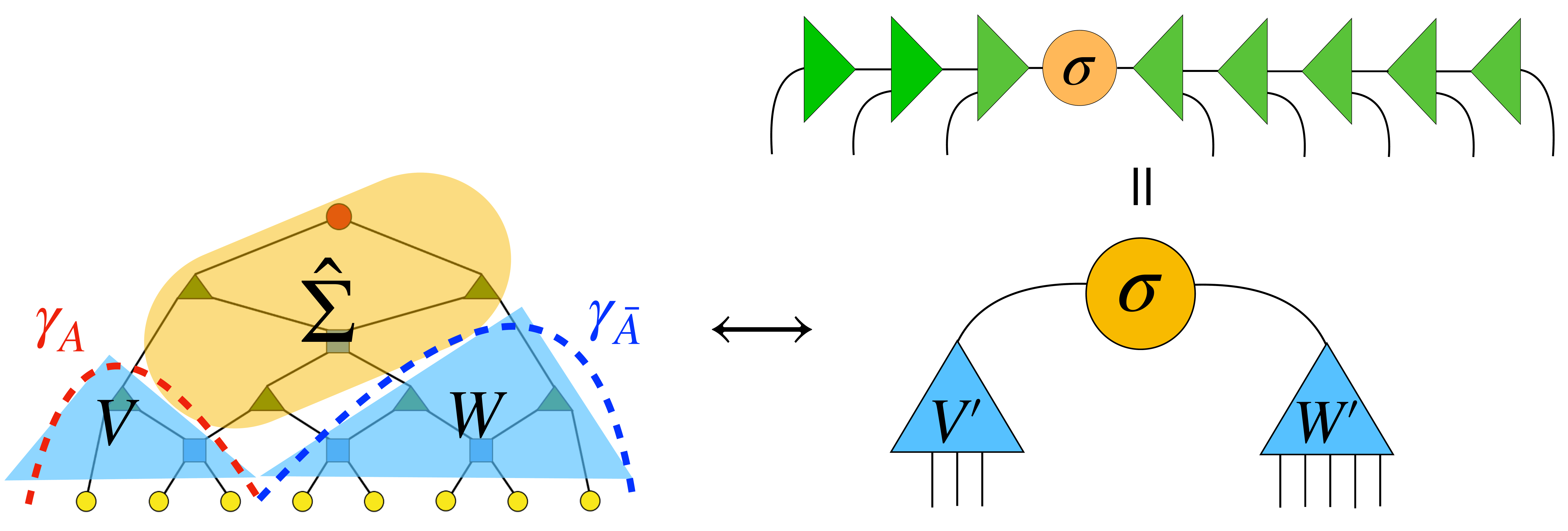}
    \caption{{MERA} can be divided into two isometries, $V$ and $W$, and the 
    {remaining} $\hat{\Sigma}${.} 
    {This structure of MERA is} in analogy with {that of} MPS, whose isometry on the left (right) of $\sigma$ is collectively denoted by $V^\prime$ ($W^\prime$). {The lower right tensor network is equivalent to a so-called one-shot entanglement distillation tensor network discussed in~\cite{Bao:2018pvs,Lin:2020ufd}.}}
    \label{fig:MERA_SVD}
\end{figure}

Fig.\ref{fig:MERA_SVD} shows a structural similarity between {MPS} {in the form} and {MERA}. The isometric parts of {the} MPS, $V^\prime,W^\prime$, correspond to {those} of {the} MERA, $V,W$, in $\mathcal{C}(A),\mathcal{C}(\bar{A})$. The singular value matrix $\sigma$ in {the} MPS corresponds to $\hat{\Sigma}$ in {the} MERA (or {the} Python's lunch in a holographic context~\cite{Brown:2019rox}). From this {view}point, 
the MPS is not only another class of tensor networks than MERA, but a simpler model sharing a common isometric structure with {the} MERA.
In the following, we will consider the MPS analogue of the entanglement distillation in {MERA} and compare the results between {the two}. 

\begin{figure}[h]
    \centering
    \includegraphics[width=0.8\linewidth]{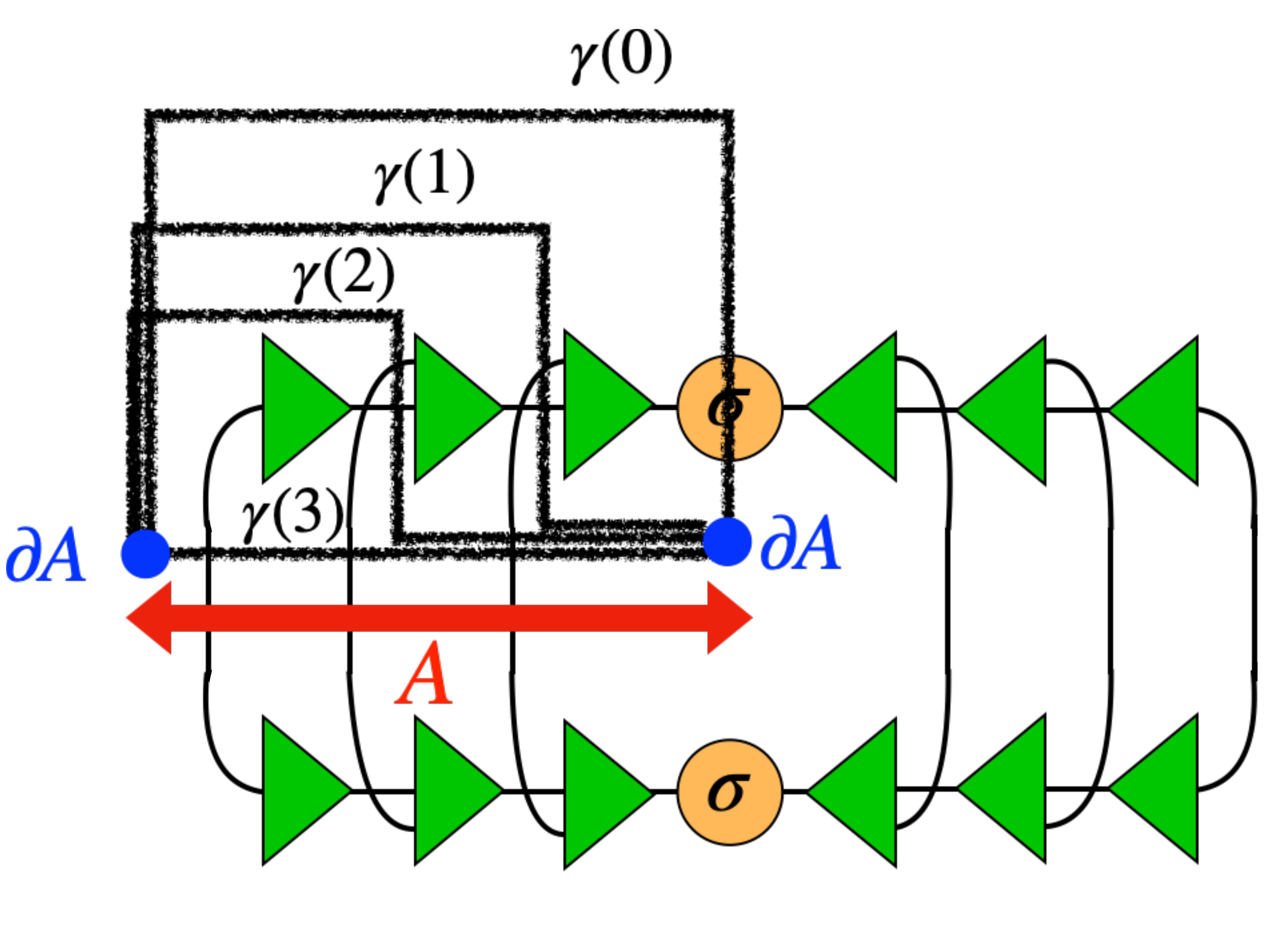}
    \caption{Foliations {interpolating the boundary subregion $A$ and the minimal bond cut surface $\gamma_A$} are denoted by $\{\gamma(\tau)\}$, where $\tau$ is an integer parametrizing each foliation.}
    \label{fig:MPS_fol}
\end{figure}

\begin{figure}[h]
    \centering
    \includegraphics[width=0.5\linewidth]{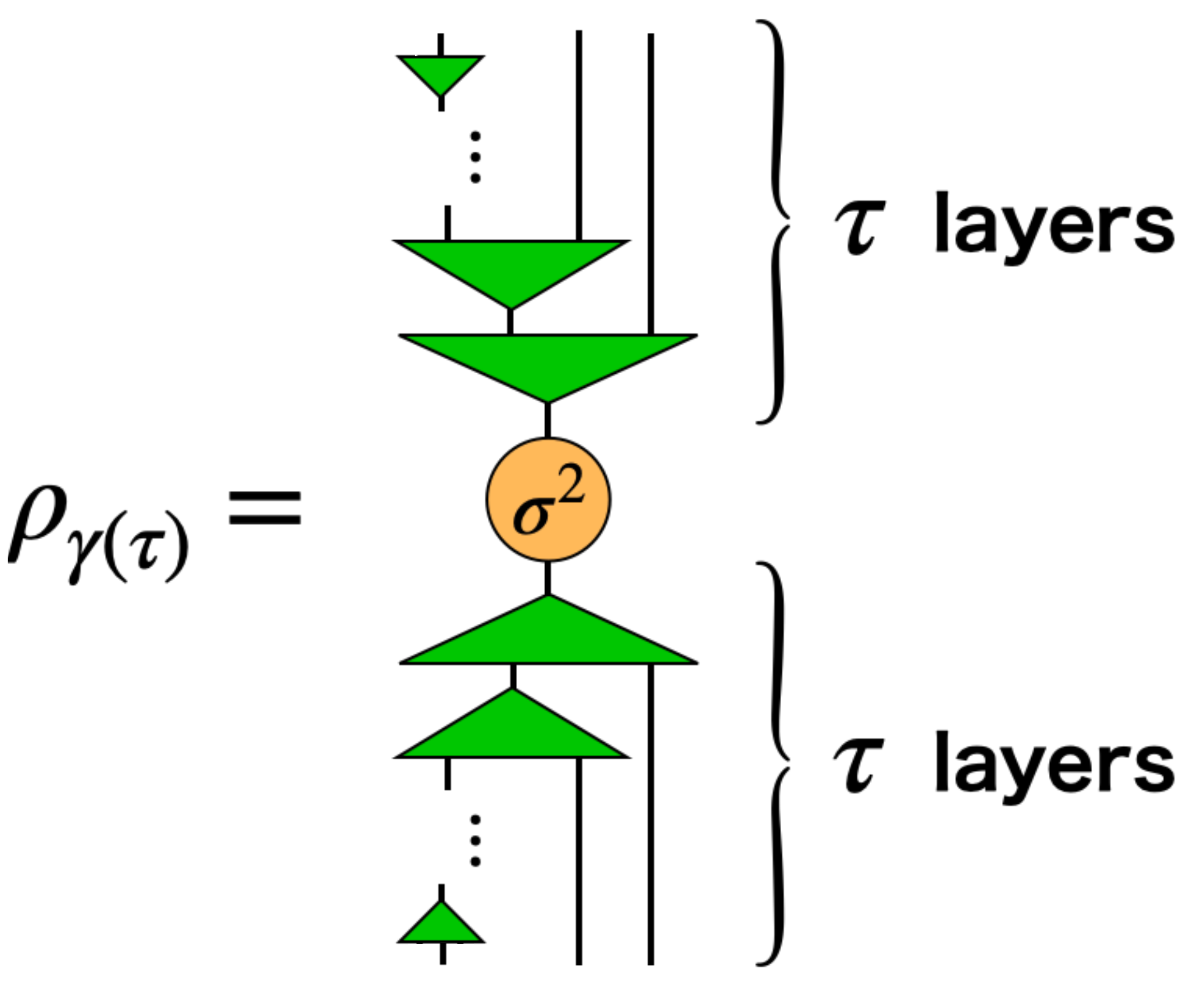}
    \caption{
    {Reduced} transition matrix associated with the foliation $\gamma(\tau)$ in Fig.\ref{fig:MPS_fol}.}
    \label{fig:MPS_red}
\end{figure}

Through the correspondence in Fig.\ref{fig:MERA_SVD}, we can consider foliations in {MPS} similar to 
{those} within $\mathcal{C}(A)$ in {MERA}. Fig.\ref{fig:MPS_fol} shows a family of foliations {$\{\gamma(\tau)\}_\tau$} in {the} MPS such that their endpoints are always fixed at the boundary of the subregion $\partial A$. Then, $\gamma_A$ can be characterized as a minimal bond cut surface, a foliation that cuts the minimum number of bonds in $\braket{\Psi}$. The foliations are chosen so that the location of the internal bond cut becomes monotonically closer to the minimal bond cut surface $\gamma_A$. The number $\tau$ specifies the number of matrices between the foliation and $\gamma_A$, parametrizing each foliation $\gamma(\tau)$. The previous discussion for the trace distance $D_\gamma$ only relies on the diagonalization of $\rho_A$ and a similarity transformation between $\rho_\gamma$. Thus, \eqref{eq:trace-dist-renyi} is 
{also} {applicable} to {MPS}. 
Given the R\'enyi-$1/2$ entropy $S_{1/2}$ of $\rho_A$, the trace distances are
\begin{align}
    D_{\gamma(\tau)}&=\sqrt{1-\frac{e^{S_{1/2}}}{\chi^{\tau+1}}},\quad \tau=0,1,2\\
    D_{\gamma(3)}&=D_{\gamma(2)}
\end{align}
since $\dim\mathcal{H}_{\gamma(\tau)}=\chi^{\tau+1}$ for $\tau=0,1,2$ and $\dim\mathcal{H}_{\gamma(3)}=\dim\mathcal{H}_{\gamma(2)}$. From this, it is apparent that the distance $D_\gamma$ decreases as the foliation approaches $\gamma_A$, i.e. $\tau$ decreases.
However, for the MPS case, it is more explicit to check {entanglement} distillation by following a state on each foliation.
The resulting reduced transition matrix on $\gamma(\tau)$ is represented by Fig.\ref{fig:MPS_red}. It can be written as 
\begin{equation}
    \rho_{\gamma{(\tau)}}= V_{\gamma{(\tau)}} \sigma^2 V_{\gamma{(\tau)}}^\dagger
\end{equation}
using an isometry $V_{\gamma{(\tau)}}$ composed of $\tau$ layers of isometries. Since only isometries act on the singular value matrix, the entanglement spectrum does not change {while the size of each $\rho_{\gamma(\tau)}$ decreases} during the distillation, which is in accordance with the necessary condition for entanglement distillation. When the foliation is a minimal bond cut surface ($\tau=0$), $V_\gamma$ becomes an identity matrix. This 
{indicates that} the distilled state via our procedure becomes diagonal $\sum_{\alpha=1}^\chi \sigma_\alpha \ket{\alpha\alpha}$ by removing isometric, redundant {degrees of freedom} from the original state.
In particular, the EPR state $\ket{\mathrm{EPR}_\chi}$ is distilled on $\gamma_A$ whenever $\sigma\propto\mathbf{1}$. Examples of such a state described by {the} MPS includes the {thermodynamic limit of the} valence-bond-solid state, i.e. the ground state of a gapped Hamiltonian called {the} AKLT model~\cite{Affleck:1987cy,Affleck:1987vf}. 

Despite the difference of criticality, the distillation in {MPS} has common features compared 
{with} that in {MERA}. In a general MERA, we discussed the monotonicity of the trace distance \eqref{eq:trace-dist-diff} toward $\gamma_\ast$. Furthermore, the distillation of the EPR state in {MERA} was equivalent to the flat entanglement spectrum \eqref{eq:schmidt}. All of these were 
shown for {the} MPS as well.

{Overall,} our distillation procedure in {MERA} can be extended to the MPS, where pushing the foliation toward the minimal bond cut surface corresponds to removing extra {degrees of freedom}, and a distillation of EPR pairs yields a flat entanglement spectrum.

\section{\label{sec:discussion}Conclusion and Discussion}
For a deeper understanding of the RT surface in holography, we have investigated how the minimal bond cut surface in tensor networks can be characterized through entanglement distillation. 
We proposed a {geometric procedure in {MERA} which is interpreted as entanglement} distillation 
and extended it to {MPS} in this paper. Cutting bonds in the norm on a surface, which we call a foliation, defines a reduced transition matrix and 
{a corresponding} state. 
To evaluate how close the state is to the EPR state, the trace distance between the EPR pair and a state on foliations has been computed. 
For a MERA, we found that the trace distance is monotonically decreasing as we push the foliation toward the minimal bond cut surface.
The numerical result also suggests that the distance was the smallest for a minimal bond cut surface.
{Compared 
{with} previous studies, our method enables us to investigate states on various bond cut surfaces even beyond a causal cone.}

For {MPS}, we extended our distillation method by properly choosing a family of foliations. Then, the boundary of the subregion is interpreted as a minimal bond cut surface. In a mixed canonical form of {the} MPS, we found that the proposed distillation process is equivalent to removing isometries, indicating a monotonic entanglement distillation toward the minimal bond cut surface. For {the} MPS, we considered changing 
the singular value distribution $\sigma$ rather than changing the bond dimension as in the random MERA. We observed when $\sigma\propto \mathbf{1}$, the EPR state can be distilled. This is a common feature both in {the} MPS and MERA. In this way, we confirmed entanglement distillation toward the minimal bond cut surface 
{in both the} MPS and MERA 
{despite} each 
{belonging} to a different criticality. The minimal bond cut surface is special in a sense that we can perform entanglement distillation toward it.

For a future direction, a numerical calculation of a larger system is desirable to relax the finite size effect. Furthermore, it is important to test our distillation procedure with variational wave functions of real ground states or {analytic solutions of tensor networks} 
like {exact MPS representations~\cite{PhysRevB.86.245305,PhysRevB.92.235150,Janik:2018kfc}} or wavelet {representations of {MERA}}~\cite{PhysRevA.97.052314,Evenbly:2016cly,Haegeman:2017vrx}. 
While the validity of our proposal should be inspected, it is intriguing whether our proposal of entanglement distillation can be understood 
from the operational meaning of each constituent tensor. {The physical interpretation of each tensor is important for a generalization to field theory as an infinitesimal transformation gives a corresponding generator for 
{entanglement distillation}.}
{In {the} case of {MERA},} 
previous studies \cite{Milsted:2018yur,Milsted:2018vop} suggest that a part of {MERA} may be interpreted as a path integral in CFT or local conformal transformations. 
{In {the} case of MPS, 
{half} of the MPS 
{acts} on the state to obtain the completely distilled state ($\rho_{\gamma(0)}$ in Fig.\ref{fig:MPS_fol}). Previous literature suggests it corresponds to a corner transfer matrix~\cite{doi:10.1143/JPSJ.66.3040,10.1007/BFb0104638} or a $\pi/2$ Euclidean modular flow~\cite{Kim:2015ygm,1999AnP...511..153P,Okunishi:2019dmv} when the MPS is prepared by transfer matrices or the Euclidean path integral.
}
As there is a formal analogy between the MPS and MERA (Fig.\ref{fig:MERA_SVD}), {we could 
{gain} insights for} entanglement distillation inside a causal cone in {MERA} 
from the analysis of {MPS}.

{
{Our} results suggest a certain geometric deformation could 
{achieve} entanglement distillation in holographic CFTs. The geometric deformation should push the boundary into the bulk. One of the candidates for such deformations is $T\overline{T}$ deformation in holographic CFTs, dual to a finite cutoff anti-de Sitter spacetime~\cite{McGough:2016lol}. Since {the} entanglement distillation we discussed preserves pseudo entropy, it also motivates us to consider a spatially inhomogeneous $T\overline{T}$ deformation as a holographic realization of pushing a foliation.} 
These observations may 
provide an alternative operational interpretation of our proposal.


{Finally, our work motivates {the development of} a novel, systematic way of constructing tensor network ansatz. We discussed how entanglement distillation arises from geometries of tensor networks. If we can reverse the procedure, a better tensor network ansatz could be constructed {on the basis of} 
algorithms of entanglement distillation~\cite{PhysRevA.97.062333,PhysRevA.64.014301,PhysRevA.64.012304,PhysRevLett.101.130502}.}

\acknowledgments
We thank 
Satoshi Iso, Shang-Ming Ruan, Ryotaro Suzuki, Tadashi Takayanagi, and Tomonori Ugajin for {their} useful discussions and comments. T. Mori also thanks 
the Quantum Information \& Foundation of Quantum Mechanics workshop for making 
{the} acquaintance of H. Manabe. Discussions during the YITP workshop YITP-W-21-19 on "Quantum Information Entropy in Physics" and T. Mori's stay at the Yukawa Institute for Theoretical Physics at Kyoto University supported by the Atoms program were useful to complete this work. H. Manabe and H. Matsueda acknowledge K. Harada for {his} discussion about tensor networks. T. Mori is supported by SOKENDAI and the Atsumi Scholarship from the Atsumi International Foundation. 
H. Matsueda is supported by KAKENHI No.21K03380, No.21H03455, No.21H04446, No.20K03769, and CSIS
in Tohoku University. 

\bibliographystyle{apsrev4-1}
\bibliography{HED}

\providecommand{\noopsort}[1]{}\providecommand{\singleletter}[1]{#1}%
\begin{thebibliography}{66}%
\makeatletter
\providecommand \@ifxundefined [1]{%
 \@ifx{#1\undefined}
}%
\providecommand \@ifnum [1]{%
 \ifnum #1\expandafter \@firstoftwo
 \else \expandafter \@secondoftwo
 \fi
}%
\providecommand \@ifx [1]{%
 \ifx #1\expandafter \@firstoftwo
 \else \expandafter \@secondoftwo
 \fi
}%
\providecommand \natexlab [1]{#1}%
\providecommand \enquote  [1]{``#1''}%
\providecommand \bibnamefont  [1]{#1}%
\providecommand \bibfnamefont [1]{#1}%
\providecommand \citenamefont [1]{#1}%
\providecommand \href@noop [0]{\@secondoftwo}%
\providecommand \href [0]{\begingroup \@sanitize@url \@href}%
\providecommand \@href[1]{\@@startlink{#1}\@@href}%
\providecommand \@@href[1]{\endgroup#1\@@endlink}%
\providecommand \@sanitize@url [0]{\catcode `\\12\catcode `\$12\catcode
  `\&12\catcode `\#12\catcode `\^12\catcode `\_12\catcode `\%12\relax}%
\providecommand \@@startlink[1]{}%
\providecommand \@@endlink[0]{}%
\providecommand \url  [0]{\begingroup\@sanitize@url \@url }%
\providecommand \@url [1]{\endgroup\@href {#1}{\urlprefix }}%
\providecommand \urlprefix  [0]{URL }%
\providecommand \Eprint [0]{\href }%
\providecommand \doibase [0]{http://dx.doi.org/}%
\providecommand \selectlanguage [0]{\@gobble}%
\providecommand \bibinfo  [0]{\@secondoftwo}%
\providecommand \bibfield  [0]{\@secondoftwo}%
\providecommand \translation [1]{[#1]}%
\providecommand \BibitemOpen [0]{}%
\providecommand \bibitemStop [0]{}%
\providecommand \bibitemNoStop [0]{.\EOS\space}%
\providecommand \EOS [0]{\spacefactor3000\relax}%
\providecommand \BibitemShut  [1]{\csname bibitem#1\endcsname}%
\let\auto@bib@innerbib\@empty
\bibitem [{\citenamefont {Susskind}(1995)}]{Susskind:1994vu}%
  \BibitemOpen
  \bibfield  {author} {\bibinfo {author} {\bibfnamefont {L.}~\bibnamefont
  {Susskind}},\ }\href {\doibase 10.1063/1.531249} {\bibfield  {journal}
  {\bibinfo  {journal} {J. Math. Phys.}\ }\textbf {\bibinfo {volume} {36}},\
  \bibinfo {pages} {6377} (\bibinfo {year} {1995})},\ \Eprint
  {http://arxiv.org/abs/hep-th/9409089} {arXiv:hep-th/9409089} \BibitemShut
  {NoStop}%
\bibitem [{\citenamefont {Maldacena}(1998)}]{Maldacena:1997re}%
  \BibitemOpen
  \bibfield  {author} {\bibinfo {author} {\bibfnamefont {J.~M.}\ \bibnamefont
  {Maldacena}},\ }\href {\doibase 10.1023/A:1026654312961} {\bibfield
  {journal} {\bibinfo  {journal} {Adv. Theor. Math. Phys.}\ }\textbf {\bibinfo
  {volume} {2}},\ \bibinfo {pages} {231} (\bibinfo {year} {1998})},\ \Eprint
  {http://arxiv.org/abs/hep-th/9711200} {arXiv:hep-th/9711200} \BibitemShut
  {NoStop}%
\bibitem [{\citenamefont {Ryu}\ and\ \citenamefont
  {Takayanagi}(2006)}]{Ryu:2006bv}%
  \BibitemOpen
  \bibfield  {author} {\bibinfo {author} {\bibfnamefont {S.}~\bibnamefont
  {Ryu}}\ and\ \bibinfo {author} {\bibfnamefont {T.}~\bibnamefont
  {Takayanagi}},\ }\href {\doibase 10.1103/PhysRevLett.96.181602} {\bibfield
  {journal} {\bibinfo  {journal} {Phys. Rev. Lett.}\ }\textbf {\bibinfo
  {volume} {96}},\ \bibinfo {pages} {181602} (\bibinfo {year} {2006})},\
  \Eprint {http://arxiv.org/abs/hep-th/0603001} {arXiv:hep-th/0603001}
  \BibitemShut {NoStop}%
\bibitem [{\citenamefont {Bekenstein}(1973)}]{Bekenstein:1973ur}%
  \BibitemOpen
  \bibfield  {author} {\bibinfo {author} {\bibfnamefont {J.~D.}\ \bibnamefont
  {Bekenstein}},\ }\href {\doibase 10.1103/PhysRevD.7.2333} {\bibfield
  {journal} {\bibinfo  {journal} {Phys. Rev. D}\ }\textbf {\bibinfo {volume}
  {7}},\ \bibinfo {pages} {2333} (\bibinfo {year} {1973})}\BibitemShut
  {NoStop}%
\bibitem [{\citenamefont {Bardeen}\ \emph {et~al.}(1973)\citenamefont
  {Bardeen}, \citenamefont {Carter},\ and\ \citenamefont
  {Hawking}}]{Bardeen:1973gs}%
  \BibitemOpen
  \bibfield  {author} {\bibinfo {author} {\bibfnamefont {J.~M.}\ \bibnamefont
  {Bardeen}}, \bibinfo {author} {\bibfnamefont {B.}~\bibnamefont {Carter}}, \
  and\ \bibinfo {author} {\bibfnamefont {S.~W.}\ \bibnamefont {Hawking}},\
  }\href {\doibase 10.1007/BF01645742} {\bibfield  {journal} {\bibinfo
  {journal} {Commun. Math. Phys.}\ }\textbf {\bibinfo {volume} {31}},\ \bibinfo
  {pages} {161} (\bibinfo {year} {1973})}\BibitemShut {NoStop}%
\bibitem [{\citenamefont {Hawking}(1975)}]{Hawking:1975vcx}%
  \BibitemOpen
  \bibfield  {author} {\bibinfo {author} {\bibfnamefont {S.~W.}\ \bibnamefont
  {Hawking}},\ }\href {\doibase 10.1007/BF02345020} {\bibfield  {journal}
  {\bibinfo  {journal} {Commun. Math. Phys.}\ }\textbf {\bibinfo {volume}
  {43}},\ \bibinfo {pages} {199} (\bibinfo {year} {1975})},\ \bibinfo {note}
  {[Erratum: Commun.Math.Phys. 46, 206 (1976)]}\BibitemShut {NoStop}%
\bibitem [{\citenamefont {Lewkowycz}\ and\ \citenamefont
  {Maldacena}(2013)}]{Lewkowycz:2013nqa}%
  \BibitemOpen
  \bibfield  {author} {\bibinfo {author} {\bibfnamefont {A.}~\bibnamefont
  {Lewkowycz}}\ and\ \bibinfo {author} {\bibfnamefont {J.}~\bibnamefont
  {Maldacena}},\ }\href {\doibase 10.1007/JHEP08(2013)090} {\bibfield
  {journal} {\bibinfo  {journal} {JHEP}\ }\textbf {\bibinfo {volume} {08}},\
  \bibinfo {pages} {090} (\bibinfo {year} {2013})},\ \Eprint
  {http://arxiv.org/abs/1304.4926} {arXiv:1304.4926 [hep-th]} \BibitemShut
  {NoStop}%
\bibitem [{\citenamefont {Bao}\ \emph {et~al.}(2019)\citenamefont {Bao},
  \citenamefont {Penington}, \citenamefont {Sorce},\ and\ \citenamefont
  {Wall}}]{Bao:2018pvs}%
  \BibitemOpen
  \bibfield  {author} {\bibinfo {author} {\bibfnamefont {N.}~\bibnamefont
  {Bao}}, \bibinfo {author} {\bibfnamefont {G.}~\bibnamefont {Penington}},
  \bibinfo {author} {\bibfnamefont {J.}~\bibnamefont {Sorce}}, \ and\ \bibinfo
  {author} {\bibfnamefont {A.~C.}\ \bibnamefont {Wall}},\ }\href {\doibase
  10.1007/JHEP11(2019)069} {\bibfield  {journal} {\bibinfo  {journal} {JHEP}\
  }\textbf {\bibinfo {volume} {11}},\ \bibinfo {pages} {069} (\bibinfo {year}
  {2019})},\ \Eprint {http://arxiv.org/abs/1812.01171} {arXiv:1812.01171
  [hep-th]} \BibitemShut {NoStop}%
\bibitem [{\citenamefont {Freedman}\ and\ \citenamefont
  {Headrick}(2017)}]{Freedman:2016zud}%
  \BibitemOpen
  \bibfield  {author} {\bibinfo {author} {\bibfnamefont {M.}~\bibnamefont
  {Freedman}}\ and\ \bibinfo {author} {\bibfnamefont {M.}~\bibnamefont
  {Headrick}},\ }\href {\doibase 10.1007/s00220-016-2796-3} {\bibfield
  {journal} {\bibinfo  {journal} {Commun. Math. Phys.}\ }\textbf {\bibinfo
  {volume} {352}},\ \bibinfo {pages} {407} (\bibinfo {year} {2017})},\ \Eprint
  {http://arxiv.org/abs/1604.00354} {arXiv:1604.00354 [hep-th]} \BibitemShut
  {NoStop}%
\bibitem [{\citenamefont {Verstraete}\ and\ \citenamefont
  {Cirac}(2004)}]{Verstraete:2004cf}%
  \BibitemOpen
  \bibfield  {author} {\bibinfo {author} {\bibfnamefont {F.}~\bibnamefont
  {Verstraete}}\ and\ \bibinfo {author} {\bibfnamefont {J.~I.}\ \bibnamefont
  {Cirac}},\ }\href@noop {} {\  (\bibinfo {year} {2004})},\ \Eprint
  {http://arxiv.org/abs/cond-mat/0407066} {arXiv:cond-mat/0407066} \BibitemShut
  {NoStop}%
\bibitem [{\citenamefont {Vidal}(2008)}]{Vidal:2008zz}%
  \BibitemOpen
  \bibfield  {author} {\bibinfo {author} {\bibfnamefont {G.}~\bibnamefont
  {Vidal}},\ }\href {\doibase 10.1103/PhysRevLett.101.110501} {\bibfield
  {journal} {\bibinfo  {journal} {Phys. Rev. Lett.}\ }\textbf {\bibinfo
  {volume} {101}},\ \bibinfo {pages} {110501} (\bibinfo {year} {2008})},\
  \Eprint {http://arxiv.org/abs/quant-ph/0610099} {arXiv:quant-ph/0610099}
  \BibitemShut {NoStop}%
\bibitem [{\citenamefont {Swingle}(2012{\natexlab{a}})}]{Swingle:2009bg}%
  \BibitemOpen
  \bibfield  {author} {\bibinfo {author} {\bibfnamefont {B.}~\bibnamefont
  {Swingle}},\ }\href {\doibase 10.1103/PhysRevD.86.065007} {\bibfield
  {journal} {\bibinfo  {journal} {Phys. Rev. D}\ }\textbf {\bibinfo {volume}
  {86}},\ \bibinfo {pages} {065007} (\bibinfo {year} {2012}{\natexlab{a}})},\
  \Eprint {http://arxiv.org/abs/0905.1317} {arXiv:0905.1317 [cond-mat.str-el]}
  \BibitemShut {NoStop}%
\bibitem [{\citenamefont {Swingle}(2012{\natexlab{b}})}]{Swingle:2012wq}%
  \BibitemOpen
  \bibfield  {author} {\bibinfo {author} {\bibfnamefont {B.}~\bibnamefont
  {Swingle}},\ }\href@noop {} {\  (\bibinfo {year} {2012}{\natexlab{b}})},\
  \Eprint {http://arxiv.org/abs/1209.3304} {arXiv:1209.3304 [hep-th]}
  \BibitemShut {NoStop}%
\bibitem [{\citenamefont {Beny}(2013)}]{Beny:2011vh}%
  \BibitemOpen
  \bibfield  {author} {\bibinfo {author} {\bibfnamefont {C.}~\bibnamefont
  {Beny}},\ }\href {\doibase 10.1088/1367-2630/15/2/023020} {\bibfield
  {journal} {\bibinfo  {journal} {New J. Phys.}\ }\textbf {\bibinfo {volume}
  {15}},\ \bibinfo {pages} {023020} (\bibinfo {year} {2013})},\ \Eprint
  {http://arxiv.org/abs/1110.4872} {arXiv:1110.4872 [quant-ph]} \BibitemShut
  {NoStop}%
\bibitem [{\citenamefont {Pastawski}\ \emph {et~al.}(2015)\citenamefont
  {Pastawski}, \citenamefont {Yoshida}, \citenamefont {Harlow},\ and\
  \citenamefont {Preskill}}]{Pastawski:2015qua}%
  \BibitemOpen
  \bibfield  {author} {\bibinfo {author} {\bibfnamefont {F.}~\bibnamefont
  {Pastawski}}, \bibinfo {author} {\bibfnamefont {B.}~\bibnamefont {Yoshida}},
  \bibinfo {author} {\bibfnamefont {D.}~\bibnamefont {Harlow}}, \ and\ \bibinfo
  {author} {\bibfnamefont {J.}~\bibnamefont {Preskill}},\ }\href {\doibase
  10.1007/JHEP06(2015)149} {\bibfield  {journal} {\bibinfo  {journal} {JHEP}\
  }\textbf {\bibinfo {volume} {06}},\ \bibinfo {pages} {149} (\bibinfo {year}
  {2015})},\ \Eprint {http://arxiv.org/abs/1503.06237} {arXiv:1503.06237
  [hep-th]} \BibitemShut {NoStop}%
\bibitem [{\citenamefont {Yang}\ \emph {et~al.}(2016)\citenamefont {Yang},
  \citenamefont {Hayden},\ and\ \citenamefont {Qi}}]{Yang:2015uoa}%
  \BibitemOpen
  \bibfield  {author} {\bibinfo {author} {\bibfnamefont {Z.}~\bibnamefont
  {Yang}}, \bibinfo {author} {\bibfnamefont {P.}~\bibnamefont {Hayden}}, \ and\
  \bibinfo {author} {\bibfnamefont {X.-L.}\ \bibnamefont {Qi}},\ }\href
  {\doibase 10.1007/JHEP01(2016)175} {\bibfield  {journal} {\bibinfo  {journal}
  {JHEP}\ }\textbf {\bibinfo {volume} {01}},\ \bibinfo {pages} {175} (\bibinfo
  {year} {2016})},\ \Eprint {http://arxiv.org/abs/1510.03784} {arXiv:1510.03784
  [hep-th]} \BibitemShut {NoStop}%
\bibitem [{\citenamefont {Czech}\ \emph {et~al.}(2016)\citenamefont {Czech},
  \citenamefont {Lamprou}, \citenamefont {McCandlish},\ and\ \citenamefont
  {Sully}}]{Czech:2015kbp}%
  \BibitemOpen
  \bibfield  {author} {\bibinfo {author} {\bibfnamefont {B.}~\bibnamefont
  {Czech}}, \bibinfo {author} {\bibfnamefont {L.}~\bibnamefont {Lamprou}},
  \bibinfo {author} {\bibfnamefont {S.}~\bibnamefont {McCandlish}}, \ and\
  \bibinfo {author} {\bibfnamefont {J.}~\bibnamefont {Sully}},\ }\href
  {\doibase 10.1007/JHEP07(2016)100} {\bibfield  {journal} {\bibinfo  {journal}
  {JHEP}\ }\textbf {\bibinfo {volume} {07}},\ \bibinfo {pages} {100} (\bibinfo
  {year} {2016})},\ \Eprint {http://arxiv.org/abs/1512.01548} {arXiv:1512.01548
  [hep-th]} \BibitemShut {NoStop}%
\bibitem [{\citenamefont {Sinai~Kunkolienkar}\ and\ \citenamefont
  {Banerjee}(2017)}]{SinaiKunkolienkar:2016lgg}%
  \BibitemOpen
  \bibfield  {author} {\bibinfo {author} {\bibfnamefont {R.}~\bibnamefont
  {Sinai~Kunkolienkar}}\ and\ \bibinfo {author} {\bibfnamefont
  {K.}~\bibnamefont {Banerjee}},\ }\href {\doibase 10.1142/S0218271817501437}
  {\bibfield  {journal} {\bibinfo  {journal} {Int. J. Mod. Phys. D}\ }\textbf
  {\bibinfo {volume} {26}},\ \bibinfo {pages} {1750143} (\bibinfo {year}
  {2017})},\ \Eprint {http://arxiv.org/abs/1611.08581} {arXiv:1611.08581
  [hep-th]} \BibitemShut {NoStop}%
\bibitem [{\citenamefont {Hayden}\ \emph {et~al.}(2016)\citenamefont {Hayden},
  \citenamefont {Nezami}, \citenamefont {Qi}, \citenamefont {Thomas},
  \citenamefont {Walter},\ and\ \citenamefont {Yang}}]{Hayden:2016cfa}%
  \BibitemOpen
  \bibfield  {author} {\bibinfo {author} {\bibfnamefont {P.}~\bibnamefont
  {Hayden}}, \bibinfo {author} {\bibfnamefont {S.}~\bibnamefont {Nezami}},
  \bibinfo {author} {\bibfnamefont {X.-L.}\ \bibnamefont {Qi}}, \bibinfo
  {author} {\bibfnamefont {N.}~\bibnamefont {Thomas}}, \bibinfo {author}
  {\bibfnamefont {M.}~\bibnamefont {Walter}}, \ and\ \bibinfo {author}
  {\bibfnamefont {Z.}~\bibnamefont {Yang}},\ }\href {\doibase
  10.1007/JHEP11(2016)009} {\bibfield  {journal} {\bibinfo  {journal} {JHEP}\
  }\textbf {\bibinfo {volume} {11}},\ \bibinfo {pages} {009} (\bibinfo {year}
  {2016})},\ \Eprint {http://arxiv.org/abs/1601.01694} {arXiv:1601.01694
  [hep-th]} \BibitemShut {NoStop}%
\bibitem [{\citenamefont {Evenbly}(2017)}]{Evenbly:2017hyg}%
  \BibitemOpen
  \bibfield  {author} {\bibinfo {author} {\bibfnamefont {G.}~\bibnamefont
  {Evenbly}},\ }\href {\doibase 10.1103/PhysRevLett.119.141602} {\bibfield
  {journal} {\bibinfo  {journal} {Phys. Rev. Lett.}\ }\textbf {\bibinfo
  {volume} {119}},\ \bibinfo {pages} {141602} (\bibinfo {year} {2017})},\
  \Eprint {http://arxiv.org/abs/1704.04229} {arXiv:1704.04229 [quant-ph]}
  \BibitemShut {NoStop}%
\bibitem [{\citenamefont {Jahn}\ \emph {et~al.}(2019)\citenamefont {Jahn},
  \citenamefont {Gluza}, \citenamefont {Pastawski},\ and\ \citenamefont
  {Eisert}}]{Jahn:2017tls}%
  \BibitemOpen
  \bibfield  {author} {\bibinfo {author} {\bibfnamefont {A.}~\bibnamefont
  {Jahn}}, \bibinfo {author} {\bibfnamefont {M.}~\bibnamefont {Gluza}},
  \bibinfo {author} {\bibfnamefont {F.}~\bibnamefont {Pastawski}}, \ and\
  \bibinfo {author} {\bibfnamefont {J.}~\bibnamefont {Eisert}},\ }\href
  {\doibase 10.1126/sciadv.aaw0092} {\bibfield  {journal} {\bibinfo  {journal}
  {Sci. Adv.}\ }\textbf {\bibinfo {volume} {5}},\ \bibinfo {pages} {eaaw0092}
  (\bibinfo {year} {2019})},\ \Eprint {http://arxiv.org/abs/1711.03109}
  {arXiv:1711.03109 [quant-ph]} \BibitemShut {NoStop}%
\bibitem [{\citenamefont {Bhattacharyya}\ \emph {et~al.}(2018)\citenamefont
  {Bhattacharyya}, \citenamefont {Hung}, \citenamefont {Lei},\ and\
  \citenamefont {Li}}]{Bhattacharyya:2017aly}%
  \BibitemOpen
  \bibfield  {author} {\bibinfo {author} {\bibfnamefont {A.}~\bibnamefont
  {Bhattacharyya}}, \bibinfo {author} {\bibfnamefont {L.-Y.}\ \bibnamefont
  {Hung}}, \bibinfo {author} {\bibfnamefont {Y.}~\bibnamefont {Lei}}, \ and\
  \bibinfo {author} {\bibfnamefont {W.}~\bibnamefont {Li}},\ }\href {\doibase
  10.1007/JHEP01(2018)139} {\bibfield  {journal} {\bibinfo  {journal} {JHEP}\
  }\textbf {\bibinfo {volume} {01}},\ \bibinfo {pages} {139} (\bibinfo {year}
  {2018})},\ \Eprint {http://arxiv.org/abs/1703.05445} {arXiv:1703.05445
  [hep-th]} \BibitemShut {NoStop}%
\bibitem [{\citenamefont {Qi}\ and\ \citenamefont {Yang}(2018)}]{Qi:2018shh}%
  \BibitemOpen
  \bibfield  {author} {\bibinfo {author} {\bibfnamefont {X.-L.}\ \bibnamefont
  {Qi}}\ and\ \bibinfo {author} {\bibfnamefont {Z.}~\bibnamefont {Yang}},\
  }\href@noop {} {\  (\bibinfo {year} {2018})},\ \Eprint
  {http://arxiv.org/abs/1801.05289} {arXiv:1801.05289 [hep-th]} \BibitemShut
  {NoStop}%
\bibitem [{\citenamefont {Milsted}\ and\ \citenamefont
  {Vidal}(2018{\natexlab{a}})}]{Milsted:2018san}%
  \BibitemOpen
  \bibfield  {author} {\bibinfo {author} {\bibfnamefont {A.}~\bibnamefont
  {Milsted}}\ and\ \bibinfo {author} {\bibfnamefont {G.}~\bibnamefont
  {Vidal}},\ }\href@noop {} {\  (\bibinfo {year} {2018}{\natexlab{a}})},\
  \Eprint {http://arxiv.org/abs/1812.00529} {arXiv:1812.00529 [hep-th]}
  \BibitemShut {NoStop}%
\bibitem [{\citenamefont {Steinberg}\ and\ \citenamefont
  {Prior}(2022)}]{Steinberg:2020bef}%
  \BibitemOpen
  \bibfield  {author} {\bibinfo {author} {\bibfnamefont {M.}~\bibnamefont
  {Steinberg}}\ and\ \bibinfo {author} {\bibfnamefont {J.}~\bibnamefont
  {Prior}},\ }\href {\doibase 10.1038/s41598-021-04375-5} {\bibfield  {journal}
  {\bibinfo  {journal} {Sci. Rep.}\ }\textbf {\bibinfo {volume} {12}},\
  \bibinfo {pages} {532} (\bibinfo {year} {2022})},\ \Eprint
  {http://arxiv.org/abs/2012.09591} {arXiv:2012.09591 [quant-ph]} \BibitemShut
  {NoStop}%
\bibitem [{\citenamefont {Jahn}\ \emph
  {et~al.}(2022{\natexlab{a}})\citenamefont {Jahn}, \citenamefont
  {Zimbor\'as},\ and\ \citenamefont {Eisert}}]{Jahn:2020ukq}%
  \BibitemOpen
  \bibfield  {author} {\bibinfo {author} {\bibfnamefont {A.}~\bibnamefont
  {Jahn}}, \bibinfo {author} {\bibfnamefont {Z.}~\bibnamefont {Zimbor\'as}}, \
  and\ \bibinfo {author} {\bibfnamefont {J.}~\bibnamefont {Eisert}},\ }\href
  {\doibase 10.22331/q-2022-02-03-643} {\bibfield  {journal} {\bibinfo
  {journal} {Quantum}\ }\textbf {\bibinfo {volume} {6}},\ \bibinfo {pages}
  {643} (\bibinfo {year} {2022}{\natexlab{a}})},\ \Eprint
  {http://arxiv.org/abs/2004.04173} {arXiv:2004.04173 [quant-ph]} \BibitemShut
  {NoStop}%
\bibitem [{\citenamefont {Jahn}\ \emph
  {et~al.}(2022{\natexlab{b}})\citenamefont {Jahn}, \citenamefont {Gluza},
  \citenamefont {Verhoeven}, \citenamefont {Singh},\ and\ \citenamefont
  {Eisert}}]{Jahn:2021kti}%
  \BibitemOpen
  \bibfield  {author} {\bibinfo {author} {\bibfnamefont {A.}~\bibnamefont
  {Jahn}}, \bibinfo {author} {\bibfnamefont {M.}~\bibnamefont {Gluza}},
  \bibinfo {author} {\bibfnamefont {C.}~\bibnamefont {Verhoeven}}, \bibinfo
  {author} {\bibfnamefont {S.}~\bibnamefont {Singh}}, \ and\ \bibinfo {author}
  {\bibfnamefont {J.}~\bibnamefont {Eisert}},\ }\href {\doibase
  10.1007/JHEP04(2022)111} {\bibfield  {journal} {\bibinfo  {journal} {JHEP}\
  }\textbf {\bibinfo {volume} {04}},\ \bibinfo {pages} {111} (\bibinfo {year}
  {2022}{\natexlab{b}})},\ \Eprint {http://arxiv.org/abs/2110.02972}
  {arXiv:2110.02972 [quant-ph]} \BibitemShut {NoStop}%
\bibitem [{\citenamefont {Ag\'on}\ \emph {et~al.}(2019)\citenamefont {Ag\'on},
  \citenamefont {De~Boer},\ and\ \citenamefont {Pedraza}}]{Agon:2018lwq}%
  \BibitemOpen
  \bibfield  {author} {\bibinfo {author} {\bibfnamefont {C.~A.}\ \bibnamefont
  {Ag\'on}}, \bibinfo {author} {\bibfnamefont {J.}~\bibnamefont {De~Boer}}, \
  and\ \bibinfo {author} {\bibfnamefont {J.~F.}\ \bibnamefont {Pedraza}},\
  }\href {\doibase 10.1007/JHEP05(2019)075} {\bibfield  {journal} {\bibinfo
  {journal} {JHEP}\ }\textbf {\bibinfo {volume} {05}},\ \bibinfo {pages} {075}
  (\bibinfo {year} {2019})},\ \Eprint {http://arxiv.org/abs/1811.08879}
  {arXiv:1811.08879 [hep-th]} \BibitemShut {NoStop}%
\bibitem [{\citenamefont {Ag\'on}\ and\ \citenamefont
  {Pedraza}(2022)}]{Agon:2021tia}%
  \BibitemOpen
  \bibfield  {author} {\bibinfo {author} {\bibfnamefont {C.~A.}\ \bibnamefont
  {Ag\'on}}\ and\ \bibinfo {author} {\bibfnamefont {J.~F.}\ \bibnamefont
  {Pedraza}},\ }\href {\doibase 10.1007/JHEP02(2022)180} {\bibfield  {journal}
  {\bibinfo  {journal} {JHEP}\ }\textbf {\bibinfo {volume} {02}},\ \bibinfo
  {pages} {180} (\bibinfo {year} {2022})},\ \Eprint
  {http://arxiv.org/abs/2105.08063} {arXiv:2105.08063 [hep-th]} \BibitemShut
  {NoStop}%
\bibitem [{\citenamefont {Rolph}(2021)}]{Rolph:2021hgz}%
  \BibitemOpen
  \bibfield  {author} {\bibinfo {author} {\bibfnamefont {A.}~\bibnamefont
  {Rolph}},\ }\href@noop {} {\  (\bibinfo {year} {2021})},\ \Eprint
  {http://arxiv.org/abs/2105.08072} {arXiv:2105.08072 [hep-th]} \BibitemShut
  {NoStop}%
\bibitem [{\citenamefont {Chen}\ \emph {et~al.}(2020)\citenamefont {Chen},
  \citenamefont {Shu},\ and\ \citenamefont {Wu}}]{Chen:2018ywy}%
  \BibitemOpen
  \bibfield  {author} {\bibinfo {author} {\bibfnamefont {C.-B.}\ \bibnamefont
  {Chen}}, \bibinfo {author} {\bibfnamefont {F.-W.}\ \bibnamefont {Shu}}, \
  and\ \bibinfo {author} {\bibfnamefont {M.-H.}\ \bibnamefont {Wu}},\ }\href
  {\doibase 10.1088/1674-1137/44/7/075102} {\bibfield  {journal} {\bibinfo
  {journal} {Chin. Phys. C}\ }\textbf {\bibinfo {volume} {44}},\ \bibinfo
  {pages} {075102} (\bibinfo {year} {2020})},\ \Eprint
  {http://arxiv.org/abs/1804.00441} {arXiv:1804.00441 [hep-th]} \BibitemShut
  {NoStop}%
\bibitem [{\citenamefont {Cui}\ \emph {et~al.}(2016)\citenamefont {Cui},
  \citenamefont {Freedman}, \citenamefont {Sattath}, \citenamefont {Stong},\
  and\ \citenamefont {Minton}}]{Cui:2015pla}%
  \BibitemOpen
  \bibfield  {author} {\bibinfo {author} {\bibfnamefont {S.~X.}\ \bibnamefont
  {Cui}}, \bibinfo {author} {\bibfnamefont {M.~H.}\ \bibnamefont {Freedman}},
  \bibinfo {author} {\bibfnamefont {O.}~\bibnamefont {Sattath}}, \bibinfo
  {author} {\bibfnamefont {R.}~\bibnamefont {Stong}}, \ and\ \bibinfo {author}
  {\bibfnamefont {G.}~\bibnamefont {Minton}},\ }\href {\doibase
  10.1063/1.4954231} {\bibfield  {journal} {\bibinfo  {journal} {J. Math.
  Phys.}\ }\textbf {\bibinfo {volume} {57}},\ \bibinfo {pages} {062206}
  (\bibinfo {year} {2016})},\ \Eprint {http://arxiv.org/abs/1508.04644}
  {arXiv:1508.04644 [math.CO]} \BibitemShut {NoStop}%
\bibitem [{\citenamefont {Lin}\ \emph {et~al.}(2021)\citenamefont {Lin},
  \citenamefont {Sun},\ and\ \citenamefont {Sun}}]{Lin:2020yzf}%
  \BibitemOpen
  \bibfield  {author} {\bibinfo {author} {\bibfnamefont {Y.-Y.}\ \bibnamefont
  {Lin}}, \bibinfo {author} {\bibfnamefont {J.-R.}\ \bibnamefont {Sun}}, \ and\
  \bibinfo {author} {\bibfnamefont {Y.}~\bibnamefont {Sun}},\ }\href {\doibase
  10.1103/PhysRevD.103.126002} {\bibfield  {journal} {\bibinfo  {journal}
  {Phys. Rev. D}\ }\textbf {\bibinfo {volume} {103}},\ \bibinfo {pages}
  {126002} (\bibinfo {year} {2021})},\ \Eprint
  {http://arxiv.org/abs/2012.05737} {arXiv:2012.05737 [hep-th]} \BibitemShut
  {NoStop}%
\bibitem [{\citenamefont {Yu}\ \emph {et~al.}(2022)\citenamefont {Yu},
  \citenamefont {Chen}, \citenamefont {Lin}, \citenamefont {Sun},\ and\
  \citenamefont {Sun}}]{Yu:2020zwk}%
  \BibitemOpen
  \bibfield  {author} {\bibinfo {author} {\bibfnamefont {C.}~\bibnamefont
  {Yu}}, \bibinfo {author} {\bibfnamefont {F.~Z.}\ \bibnamefont {Chen}},
  \bibinfo {author} {\bibfnamefont {Y.-Y.}\ \bibnamefont {Lin}}, \bibinfo
  {author} {\bibfnamefont {J.-R.}\ \bibnamefont {Sun}}, \ and\ \bibinfo
  {author} {\bibfnamefont {Y.}~\bibnamefont {Sun}},\ }\href {\doibase
  10.1088/1674-1137/ac69ba} {\bibfield  {journal} {\bibinfo  {journal} {Chin.
  Phys. C}\ }\textbf {\bibinfo {volume} {46}},\ \bibinfo {pages} {085104}
  (\bibinfo {year} {2022})},\ \Eprint {http://arxiv.org/abs/2010.03167}
  {arXiv:2010.03167 [hep-th]} \BibitemShut {NoStop}%
\bibitem [{\citenamefont {Lin}\ \emph {et~al.}(2020)\citenamefont {Lin},
  \citenamefont {Sun},\ and\ \citenamefont {Sun}}]{Lin:2020ufd}%
  \BibitemOpen
  \bibfield  {author} {\bibinfo {author} {\bibfnamefont {Y.-Y.}\ \bibnamefont
  {Lin}}, \bibinfo {author} {\bibfnamefont {J.-R.}\ \bibnamefont {Sun}}, \ and\
  \bibinfo {author} {\bibfnamefont {Y.}~\bibnamefont {Sun}},\ }\href {\doibase
  10.1007/JHEP12(2020)083} {\bibfield  {journal} {\bibinfo  {journal} {JHEP}\
  }\textbf {\bibinfo {volume} {12}},\ \bibinfo {pages} {083} (\bibinfo {year}
  {2020})},\ \Eprint {http://arxiv.org/abs/2010.01907} {arXiv:2010.01907
  [hep-th]} \BibitemShut {NoStop}%
\bibitem [{\citenamefont {Evenbly}\ and\ \citenamefont
  {Vidal}(2009)}]{Evenbly:2007hxg}%
  \BibitemOpen
  \bibfield  {author} {\bibinfo {author} {\bibfnamefont {G.}~\bibnamefont
  {Evenbly}}\ and\ \bibinfo {author} {\bibfnamefont {G.}~\bibnamefont
  {Vidal}},\ }\href {\doibase 10.1103/PhysRevB.79.144108} {\bibfield  {journal}
  {\bibinfo  {journal} {Phys. Rev. B}\ }\textbf {\bibinfo {volume} {79}},\
  \bibinfo {pages} {144108} (\bibinfo {year} {2009})},\ \Eprint
  {http://arxiv.org/abs/0707.1454} {arXiv:0707.1454 [cond-mat.str-el]}
  \BibitemShut {NoStop}%
\bibitem [{\citenamefont {Evenbly}\ and\ \citenamefont
  {Vidal}(2013)}]{evenbly2013quantum}%
  \BibitemOpen
  \bibfield  {author} {\bibinfo {author} {\bibfnamefont {G.}~\bibnamefont
  {Evenbly}}\ and\ \bibinfo {author} {\bibfnamefont {G.}~\bibnamefont
  {Vidal}},\ }\href@noop {} {\enquote {\bibinfo {title} {Quantum criticality
  with the multi-scale entanglement renormalization ansatz},}\ } (\bibinfo
  {year} {2013}),\ \Eprint {http://arxiv.org/abs/1109.5334} {arXiv:1109.5334
  [quant-ph]} \BibitemShut {NoStop}%
\bibitem [{\citenamefont {Nakata}\ \emph {et~al.}(2021)\citenamefont {Nakata},
  \citenamefont {Takayanagi}, \citenamefont {Taki}, \citenamefont {Tamaoka},\
  and\ \citenamefont {Wei}}]{Nakata:2020luh}%
  \BibitemOpen
  \bibfield  {author} {\bibinfo {author} {\bibfnamefont {Y.}~\bibnamefont
  {Nakata}}, \bibinfo {author} {\bibfnamefont {T.}~\bibnamefont {Takayanagi}},
  \bibinfo {author} {\bibfnamefont {Y.}~\bibnamefont {Taki}}, \bibinfo {author}
  {\bibfnamefont {K.}~\bibnamefont {Tamaoka}}, \ and\ \bibinfo {author}
  {\bibfnamefont {Z.}~\bibnamefont {Wei}},\ }\href {\doibase
  10.1103/PhysRevD.103.026005} {\bibfield  {journal} {\bibinfo  {journal}
  {Phys. Rev. D}\ }\textbf {\bibinfo {volume} {103}},\ \bibinfo {pages}
  {026005} (\bibinfo {year} {2021})},\ \Eprint
  {http://arxiv.org/abs/2005.13801} {arXiv:2005.13801 [hep-th]} \BibitemShut
  {NoStop}%
\bibitem [{\citenamefont {Bertini}\ \emph
  {et~al.}(2019{\natexlab{a}})\citenamefont {Bertini}, \citenamefont {Kos},\
  and\ \citenamefont {Prosen}}]{Bertini:2018fbz}%
  \BibitemOpen
  \bibfield  {author} {\bibinfo {author} {\bibfnamefont {B.}~\bibnamefont
  {Bertini}}, \bibinfo {author} {\bibfnamefont {P.}~\bibnamefont {Kos}}, \ and\
  \bibinfo {author} {\bibfnamefont {T.}~\bibnamefont {Prosen}},\ }\href
  {\doibase 10.1103/PhysRevX.9.021033} {\bibfield  {journal} {\bibinfo
  {journal} {Phys. Rev. X}\ }\textbf {\bibinfo {volume} {9}},\ \bibinfo {pages}
  {021033} (\bibinfo {year} {2019}{\natexlab{a}})},\ \Eprint
  {http://arxiv.org/abs/1812.05090} {arXiv:1812.05090 [cond-mat.stat-mech]}
  \BibitemShut {NoStop}%
\bibitem [{\citenamefont {Bertini}\ \emph
  {et~al.}(2019{\natexlab{b}})\citenamefont {Bertini}, \citenamefont {Kos},\
  and\ \citenamefont {Prosen}}]{Bertini:2019lgy}%
  \BibitemOpen
  \bibfield  {author} {\bibinfo {author} {\bibfnamefont {B.}~\bibnamefont
  {Bertini}}, \bibinfo {author} {\bibfnamefont {P.}~\bibnamefont {Kos}}, \ and\
  \bibinfo {author} {\bibfnamefont {T.}~\bibnamefont {Prosen}},\ }\href
  {\doibase 10.1103/PhysRevLett.123.210601} {\bibfield  {journal} {\bibinfo
  {journal} {Phys. Rev. Lett.}\ }\textbf {\bibinfo {volume} {123}},\ \bibinfo
  {pages} {210601} (\bibinfo {year} {2019}{\natexlab{b}})},\ \Eprint
  {http://arxiv.org/abs/1904.02140} {arXiv:1904.02140 [cond-mat.stat-mech]}
  \BibitemShut {NoStop}%
\bibitem [{\citenamefont {Lewkowycz}\ \emph {et~al.}(2020)\citenamefont
  {Lewkowycz}, \citenamefont {Liu}, \citenamefont {Silverstein},\ and\
  \citenamefont {Torroba}}]{Lewkowycz:2019xse}%
  \BibitemOpen
  \bibfield  {author} {\bibinfo {author} {\bibfnamefont {A.}~\bibnamefont
  {Lewkowycz}}, \bibinfo {author} {\bibfnamefont {J.}~\bibnamefont {Liu}},
  \bibinfo {author} {\bibfnamefont {E.}~\bibnamefont {Silverstein}}, \ and\
  \bibinfo {author} {\bibfnamefont {G.}~\bibnamefont {Torroba}},\ }\href
  {\doibase 10.1007/JHEP04(2020)152} {\bibfield  {journal} {\bibinfo  {journal}
  {JHEP}\ }\textbf {\bibinfo {volume} {04}},\ \bibinfo {pages} {152} (\bibinfo
  {year} {2020})},\ \Eprint {http://arxiv.org/abs/1909.13808} {arXiv:1909.13808
  [hep-th]} \BibitemShut {NoStop}%
\bibitem [{\citenamefont {Kudler-Flam}\ \emph {et~al.}(2020)\citenamefont
  {Kudler-Flam}, \citenamefont {Nozaki}, \citenamefont {Ryu},\ and\
  \citenamefont {Tan}}]{Kudler-Flam:2019wtv}%
  \BibitemOpen
  \bibfield  {author} {\bibinfo {author} {\bibfnamefont {J.}~\bibnamefont
  {Kudler-Flam}}, \bibinfo {author} {\bibfnamefont {M.}~\bibnamefont {Nozaki}},
  \bibinfo {author} {\bibfnamefont {S.}~\bibnamefont {Ryu}}, \ and\ \bibinfo
  {author} {\bibfnamefont {M.~T.}\ \bibnamefont {Tan}},\ }\href {\doibase
  10.1007/JHEP01(2020)031} {\bibfield  {journal} {\bibinfo  {journal} {JHEP}\
  }\textbf {\bibinfo {volume} {01}},\ \bibinfo {pages} {031} (\bibinfo {year}
  {2020})},\ \Eprint {http://arxiv.org/abs/1906.07639} {arXiv:1906.07639
  [hep-th]} \BibitemShut {NoStop}%
\bibitem [{\citenamefont {Gray}(2018)}]{Gray2018}%
  \BibitemOpen
  \bibfield  {author} {\bibinfo {author} {\bibfnamefont {J.}~\bibnamefont
  {Gray}},\ }\href {\doibase 10.21105/joss.00819} {\bibfield  {journal}
  {\bibinfo  {journal} {Journal of Open Source Software}\ }\textbf {\bibinfo
  {volume} {3}},\ \bibinfo {pages} {819} (\bibinfo {year} {2018})}\BibitemShut
  {NoStop}%
\bibitem [{\citenamefont {Gray}\ and\ \citenamefont
  {Kourtis}(2021)}]{Gray2021}%
  \BibitemOpen
  \bibfield  {author} {\bibinfo {author} {\bibfnamefont {J.}~\bibnamefont
  {Gray}}\ and\ \bibinfo {author} {\bibfnamefont {S.}~\bibnamefont {Kourtis}},\
  }\href {\doibase 10.22331/q-2021-03-15-410} {\bibfield  {journal} {\bibinfo
  {journal} {Quantum}\ }\textbf {\bibinfo {volume} {5}},\ \bibinfo {pages}
  {410} (\bibinfo {year} {2021})}\BibitemShut {NoStop}%
\bibitem [{\citenamefont {Schollwöck}(2011)}]{SCHOLLWOCK201196}%
  \BibitemOpen
  \bibfield  {author} {\bibinfo {author} {\bibfnamefont {U.}~\bibnamefont
  {Schollwöck}},\ }\href {\doibase https://doi.org/10.1016/j.aop.2010.09.012}
  {\bibfield  {journal} {\bibinfo  {journal} {Annals of Physics}\ }\textbf
  {\bibinfo {volume} {326}},\ \bibinfo {pages} {96} (\bibinfo {year} {2011})},\
  \bibinfo {note} {january 2011 Special Issue}\BibitemShut {NoStop}%
\bibitem [{\citenamefont {Brown}\ \emph {et~al.}(2020)\citenamefont {Brown},
  \citenamefont {Gharibyan}, \citenamefont {Penington},\ and\ \citenamefont
  {Susskind}}]{Brown:2019rox}%
  \BibitemOpen
  \bibfield  {author} {\bibinfo {author} {\bibfnamefont {A.~R.}\ \bibnamefont
  {Brown}}, \bibinfo {author} {\bibfnamefont {H.}~\bibnamefont {Gharibyan}},
  \bibinfo {author} {\bibfnamefont {G.}~\bibnamefont {Penington}}, \ and\
  \bibinfo {author} {\bibfnamefont {L.}~\bibnamefont {Susskind}},\ }\href
  {\doibase 10.1007/JHEP08(2020)121} {\bibfield  {journal} {\bibinfo  {journal}
  {JHEP}\ }\textbf {\bibinfo {volume} {08}},\ \bibinfo {pages} {121} (\bibinfo
  {year} {2020})},\ \Eprint {http://arxiv.org/abs/1912.00228} {arXiv:1912.00228
  [hep-th]} \BibitemShut {NoStop}%
\bibitem [{\citenamefont {Affleck}\ \emph {et~al.}(1988)\citenamefont
  {Affleck}, \citenamefont {Kennedy}, \citenamefont {Lieb},\ and\ \citenamefont
  {Tasaki}}]{Affleck:1987cy}%
  \BibitemOpen
  \bibfield  {author} {\bibinfo {author} {\bibfnamefont {I.}~\bibnamefont
  {Affleck}}, \bibinfo {author} {\bibfnamefont {T.}~\bibnamefont {Kennedy}},
  \bibinfo {author} {\bibfnamefont {E.~H.}\ \bibnamefont {Lieb}}, \ and\
  \bibinfo {author} {\bibfnamefont {H.}~\bibnamefont {Tasaki}},\ }\href
  {\doibase 10.1007/BF01218021} {\bibfield  {journal} {\bibinfo  {journal}
  {Commun. Math. Phys.}\ }\textbf {\bibinfo {volume} {115}},\ \bibinfo {pages}
  {477} (\bibinfo {year} {1988})}\BibitemShut {NoStop}%
\bibitem [{\citenamefont {Affleck}\ \emph {et~al.}(1987)\citenamefont
  {Affleck}, \citenamefont {Kennedy}, \citenamefont {Lieb},\ and\ \citenamefont
  {Tasaki}}]{Affleck:1987vf}%
  \BibitemOpen
  \bibfield  {author} {\bibinfo {author} {\bibfnamefont {I.}~\bibnamefont
  {Affleck}}, \bibinfo {author} {\bibfnamefont {T.}~\bibnamefont {Kennedy}},
  \bibinfo {author} {\bibfnamefont {E.~H.}\ \bibnamefont {Lieb}}, \ and\
  \bibinfo {author} {\bibfnamefont {H.}~\bibnamefont {Tasaki}},\ }\href
  {\doibase 10.1103/PhysRevLett.59.799} {\bibfield  {journal} {\bibinfo
  {journal} {Phys. Rev. Lett.}\ }\textbf {\bibinfo {volume} {59}},\ \bibinfo
  {pages} {799} (\bibinfo {year} {1987})}\BibitemShut {NoStop}%
\bibitem [{\citenamefont {Zaletel}\ and\ \citenamefont
  {Mong}(2012)}]{PhysRevB.86.245305}%
  \BibitemOpen
  \bibfield  {author} {\bibinfo {author} {\bibfnamefont {M.~P.}\ \bibnamefont
  {Zaletel}}\ and\ \bibinfo {author} {\bibfnamefont {R.~S.~K.}\ \bibnamefont
  {Mong}},\ }\href {\doibase 10.1103/PhysRevB.86.245305} {\bibfield  {journal}
  {\bibinfo  {journal} {Phys. Rev. B}\ }\textbf {\bibinfo {volume} {86}},\
  \bibinfo {pages} {245305} (\bibinfo {year} {2012})}\BibitemShut {NoStop}%
\bibitem [{\citenamefont {Rams}\ \emph {et~al.}(2015)\citenamefont {Rams},
  \citenamefont {Zauner}, \citenamefont {Bal}, \citenamefont {Haegeman},\ and\
  \citenamefont {Verstraete}}]{PhysRevB.92.235150}%
  \BibitemOpen
  \bibfield  {author} {\bibinfo {author} {\bibfnamefont {M.~M.}\ \bibnamefont
  {Rams}}, \bibinfo {author} {\bibfnamefont {V.}~\bibnamefont {Zauner}},
  \bibinfo {author} {\bibfnamefont {M.}~\bibnamefont {Bal}}, \bibinfo {author}
  {\bibfnamefont {J.}~\bibnamefont {Haegeman}}, \ and\ \bibinfo {author}
  {\bibfnamefont {F.}~\bibnamefont {Verstraete}},\ }\href {\doibase
  10.1103/PhysRevB.92.235150} {\bibfield  {journal} {\bibinfo  {journal} {Phys.
  Rev. B}\ }\textbf {\bibinfo {volume} {92}},\ \bibinfo {pages} {235150}
  (\bibinfo {year} {2015})}\BibitemShut {NoStop}%
\bibitem [{\citenamefont {Janik}(2019)}]{Janik:2018kfc}%
  \BibitemOpen
  \bibfield  {author} {\bibinfo {author} {\bibfnamefont {R.~A.}\ \bibnamefont
  {Janik}},\ }\href {\doibase 10.1007/JHEP01(2019)225} {\bibfield  {journal}
  {\bibinfo  {journal} {JHEP}\ }\textbf {\bibinfo {volume} {01}},\ \bibinfo
  {pages} {225} (\bibinfo {year} {2019})},\ \Eprint
  {http://arxiv.org/abs/1811.11027} {arXiv:1811.11027 [hep-th]} \BibitemShut
  {NoStop}%
\bibitem [{\citenamefont {Evenbly}\ and\ \citenamefont
  {White}(2018)}]{PhysRevA.97.052314}%
  \BibitemOpen
  \bibfield  {author} {\bibinfo {author} {\bibfnamefont {G.}~\bibnamefont
  {Evenbly}}\ and\ \bibinfo {author} {\bibfnamefont {S.~R.}\ \bibnamefont
  {White}},\ }\href {\doibase 10.1103/PhysRevA.97.052314} {\bibfield  {journal}
  {\bibinfo  {journal} {Phys. Rev. A}\ }\textbf {\bibinfo {volume} {97}},\
  \bibinfo {pages} {052314} (\bibinfo {year} {2018})}\BibitemShut {NoStop}%
\bibitem [{\citenamefont {Evenbly}\ and\ \citenamefont
  {White}(2016)}]{Evenbly:2016cly}%
  \BibitemOpen
  \bibfield  {author} {\bibinfo {author} {\bibfnamefont {G.}~\bibnamefont
  {Evenbly}}\ and\ \bibinfo {author} {\bibfnamefont {S.~R.}\ \bibnamefont
  {White}},\ }\href {\doibase 10.1103/PhysRevLett.116.140403} {\bibfield
  {journal} {\bibinfo  {journal} {Phys. Rev. Lett.}\ }\textbf {\bibinfo
  {volume} {116}},\ \bibinfo {pages} {140403} (\bibinfo {year} {2016})},\
  \Eprint {http://arxiv.org/abs/1602.01166} {arXiv:1602.01166
  [cond-mat.str-el]} \BibitemShut {NoStop}%
\bibitem [{\citenamefont {Haegeman}\ \emph {et~al.}(2018)\citenamefont
  {Haegeman}, \citenamefont {Swingle}, \citenamefont {Walter}, \citenamefont
  {Cotler}, \citenamefont {Evenbly},\ and\ \citenamefont
  {Scholz}}]{Haegeman:2017vrx}%
  \BibitemOpen
  \bibfield  {author} {\bibinfo {author} {\bibfnamefont {J.}~\bibnamefont
  {Haegeman}}, \bibinfo {author} {\bibfnamefont {B.}~\bibnamefont {Swingle}},
  \bibinfo {author} {\bibfnamefont {M.}~\bibnamefont {Walter}}, \bibinfo
  {author} {\bibfnamefont {J.}~\bibnamefont {Cotler}}, \bibinfo {author}
  {\bibfnamefont {G.}~\bibnamefont {Evenbly}}, \ and\ \bibinfo {author}
  {\bibfnamefont {V.~B.}\ \bibnamefont {Scholz}},\ }\href {\doibase
  10.1103/PhysRevX.8.011003} {\bibfield  {journal} {\bibinfo  {journal} {Phys.
  Rev. X}\ }\textbf {\bibinfo {volume} {8}},\ \bibinfo {pages} {011003}
  (\bibinfo {year} {2018})},\ \Eprint {http://arxiv.org/abs/1707.06243}
  {arXiv:1707.06243 [quant-ph]} \BibitemShut {NoStop}%
\bibitem [{\citenamefont {Milsted}\ and\ \citenamefont
  {Vidal}(2018{\natexlab{b}})}]{Milsted:2018yur}%
  \BibitemOpen
  \bibfield  {author} {\bibinfo {author} {\bibfnamefont {A.}~\bibnamefont
  {Milsted}}\ and\ \bibinfo {author} {\bibfnamefont {G.}~\bibnamefont
  {Vidal}},\ }\href@noop {} {\  (\bibinfo {year} {2018}{\natexlab{b}})},\
  \Eprint {http://arxiv.org/abs/1807.02501} {arXiv:1807.02501
  [cond-mat.str-el]} \BibitemShut {NoStop}%
\bibitem [{\citenamefont {Milsted}\ and\ \citenamefont
  {Vidal}(2018{\natexlab{c}})}]{Milsted:2018vop}%
  \BibitemOpen
  \bibfield  {author} {\bibinfo {author} {\bibfnamefont {A.}~\bibnamefont
  {Milsted}}\ and\ \bibinfo {author} {\bibfnamefont {G.}~\bibnamefont
  {Vidal}},\ }\href@noop {} {\  (\bibinfo {year} {2018}{\natexlab{c}})},\
  \Eprint {http://arxiv.org/abs/1805.12524} {arXiv:1805.12524
  [cond-mat.str-el]} \BibitemShut {NoStop}%
\bibitem [{\citenamefont {Nishino}\ and\ \citenamefont
  {Okunishi}(1997{\natexlab{a}})}]{doi:10.1143/JPSJ.66.3040}%
  \BibitemOpen
  \bibfield  {author} {\bibinfo {author} {\bibfnamefont {T.}~\bibnamefont
  {Nishino}}\ and\ \bibinfo {author} {\bibfnamefont {K.}~\bibnamefont
  {Okunishi}},\ }\href {\doibase 10.1143/JPSJ.66.3040} {\bibfield  {journal}
  {\bibinfo  {journal} {Journal of the Physical Society of Japan}\ }\textbf
  {\bibinfo {volume} {66}},\ \bibinfo {pages} {3040} (\bibinfo {year}
  {1997}{\natexlab{a}})},\ \Eprint
  {http://arxiv.org/abs/https://doi.org/10.1143/JPSJ.66.3040}
  {https://doi.org/10.1143/JPSJ.66.3040} \BibitemShut {NoStop}%
\bibitem [{\citenamefont {Nishino}\ and\ \citenamefont
  {Okunishi}(1997{\natexlab{b}})}]{10.1007/BFb0104638}%
  \BibitemOpen
  \bibfield  {author} {\bibinfo {author} {\bibfnamefont {T.}~\bibnamefont
  {Nishino}}\ and\ \bibinfo {author} {\bibfnamefont {K.}~\bibnamefont
  {Okunishi}},\ }in\ \href@noop {} {\emph {\bibinfo {booktitle} {Strongly
  Correlated Magnetic and Superconducting Systems}}},\ \bibinfo {editor}
  {edited by\ \bibinfo {editor} {\bibfnamefont {G.}~\bibnamefont {Sierra}}\
  and\ \bibinfo {editor} {\bibfnamefont {M.~A.}\ \bibnamefont
  {Mart{\'i}n-Delgado}}}\ (\bibinfo  {publisher} {Springer Berlin Heidelberg},\
  \bibinfo {address} {Berlin, Heidelberg},\ \bibinfo {year} {1997})\ pp.\
  \bibinfo {pages} {167--183}\BibitemShut {NoStop}%
\bibitem [{\citenamefont {Kim}\ \emph {et~al.}(2016)\citenamefont {Kim},
  \citenamefont {Katsura}, \citenamefont {Trivedi},\ and\ \citenamefont
  {Han}}]{Kim:2015ygm}%
  \BibitemOpen
  \bibfield  {author} {\bibinfo {author} {\bibfnamefont {P.}~\bibnamefont
  {Kim}}, \bibinfo {author} {\bibfnamefont {H.}~\bibnamefont {Katsura}},
  \bibinfo {author} {\bibfnamefont {N.}~\bibnamefont {Trivedi}}, \ and\
  \bibinfo {author} {\bibfnamefont {J.~H.}\ \bibnamefont {Han}},\ }\href
  {\doibase 10.1103/PhysRevB.94.195110} {\bibfield  {journal} {\bibinfo
  {journal} {Phys. Rev. B}\ }\textbf {\bibinfo {volume} {94}},\ \bibinfo
  {pages} {195110} (\bibinfo {year} {2016})},\ \Eprint
  {http://arxiv.org/abs/1512.08597} {arXiv:1512.08597 [cond-mat.str-el]}
  \BibitemShut {NoStop}%
\bibitem [{\citenamefont {{Peschel}}\ \emph {et~al.}(1999)\citenamefont
  {{Peschel}}, \citenamefont {{Kaulke}},\ and\ \citenamefont
  {{Legeza}}}]{1999AnP...511..153P}%
  \BibitemOpen
  \bibfield  {author} {\bibinfo {author} {\bibfnamefont {I.}~\bibnamefont
  {{Peschel}}}, \bibinfo {author} {\bibfnamefont {M.}~\bibnamefont {{Kaulke}}},
  \ and\ \bibinfo {author} {\bibfnamefont {{\"O}.}~\bibnamefont {{Legeza}}},\
  }\href {\doibase
  10.1002/(SICI)1521-3889(199902)8:2\textless{}153::AID-ANDP153\textgreater{}3.0.CO;2-N}
  {\bibfield  {journal} {\bibinfo  {journal} {Annalen der Physik}\ }\textbf
  {\bibinfo {volume} {8}},\ \bibinfo {pages} {153} (\bibinfo {year} {1999})},\
  \Eprint {http://arxiv.org/abs/cond-mat/9810174} {arXiv:cond-mat/9810174
  [cond-mat.stat-mech]} \BibitemShut {NoStop}%
\bibitem [{\citenamefont {Okunishi}\ and\ \citenamefont
  {Seki}(2019)}]{Okunishi:2019dmv}%
  \BibitemOpen
  \bibfield  {author} {\bibinfo {author} {\bibfnamefont {K.}~\bibnamefont
  {Okunishi}}\ and\ \bibinfo {author} {\bibfnamefont {K.}~\bibnamefont
  {Seki}},\ }\href {\doibase 10.7566/JPSJ.88.114002} {\bibfield  {journal}
  {\bibinfo  {journal} {J. Phys. Soc. Jap.}\ }\textbf {\bibinfo {volume}
  {88}},\ \bibinfo {pages} {114002} (\bibinfo {year} {2019})},\ \Eprint
  {http://arxiv.org/abs/1906.10441} {arXiv:1906.10441 [cond-mat.quant-gas]}
  \BibitemShut {NoStop}%
\bibitem [{\citenamefont {McGough}\ \emph {et~al.}(2018)\citenamefont
  {McGough}, \citenamefont {Mezei},\ and\ \citenamefont
  {Verlinde}}]{McGough:2016lol}%
  \BibitemOpen
  \bibfield  {author} {\bibinfo {author} {\bibfnamefont {L.}~\bibnamefont
  {McGough}}, \bibinfo {author} {\bibfnamefont {M.}~\bibnamefont {Mezei}}, \
  and\ \bibinfo {author} {\bibfnamefont {H.}~\bibnamefont {Verlinde}},\ }\href
  {\doibase 10.1007/JHEP04(2018)010} {\bibfield  {journal} {\bibinfo  {journal}
  {JHEP}\ }\textbf {\bibinfo {volume} {04}},\ \bibinfo {pages} {010} (\bibinfo
  {year} {2018})},\ \Eprint {http://arxiv.org/abs/1611.03470} {arXiv:1611.03470
  [hep-th]} \BibitemShut {NoStop}%
\bibitem [{\citenamefont {Rozp\ifmmode~\mbox{\k{e}}\else \k{e}\fi{}dek}\ \emph
  {et~al.}(2018)\citenamefont {Rozp\ifmmode~\mbox{\k{e}}\else \k{e}\fi{}dek},
  \citenamefont {Schiet}, \citenamefont {Thinh}, \citenamefont {Elkouss},
  \citenamefont {Doherty},\ and\ \citenamefont {Wehner}}]{PhysRevA.97.062333}%
  \BibitemOpen
  \bibfield  {author} {\bibinfo {author} {\bibfnamefont {F.}~\bibnamefont
  {Rozp\ifmmode~\mbox{\k{e}}\else \k{e}\fi{}dek}}, \bibinfo {author}
  {\bibfnamefont {T.}~\bibnamefont {Schiet}}, \bibinfo {author} {\bibfnamefont
  {L.~P.}\ \bibnamefont {Thinh}}, \bibinfo {author} {\bibfnamefont
  {D.}~\bibnamefont {Elkouss}}, \bibinfo {author} {\bibfnamefont {A.~C.}\
  \bibnamefont {Doherty}}, \ and\ \bibinfo {author} {\bibfnamefont
  {S.}~\bibnamefont {Wehner}},\ }\href {\doibase 10.1103/PhysRevA.97.062333}
  {\bibfield  {journal} {\bibinfo  {journal} {Phys. Rev. A}\ }\textbf {\bibinfo
  {volume} {97}},\ \bibinfo {pages} {062333} (\bibinfo {year}
  {2018})}\BibitemShut {NoStop}%
\bibitem [{\citenamefont {Zhao}\ \emph {et~al.}(2001)\citenamefont {Zhao},
  \citenamefont {Pan},\ and\ \citenamefont {Zhan}}]{PhysRevA.64.014301}%
  \BibitemOpen
  \bibfield  {author} {\bibinfo {author} {\bibfnamefont {Z.}~\bibnamefont
  {Zhao}}, \bibinfo {author} {\bibfnamefont {J.-W.}\ \bibnamefont {Pan}}, \
  and\ \bibinfo {author} {\bibfnamefont {M.~S.}\ \bibnamefont {Zhan}},\ }\href
  {\doibase 10.1103/PhysRevA.64.014301} {\bibfield  {journal} {\bibinfo
  {journal} {Phys. Rev. A}\ }\textbf {\bibinfo {volume} {64}},\ \bibinfo
  {pages} {014301} (\bibinfo {year} {2001})}\BibitemShut {NoStop}%
\bibitem [{\citenamefont {Yamamoto}\ \emph {et~al.}(2001)\citenamefont
  {Yamamoto}, \citenamefont {Koashi},\ and\ \citenamefont
  {Imoto}}]{PhysRevA.64.012304}%
  \BibitemOpen
  \bibfield  {author} {\bibinfo {author} {\bibfnamefont {T.}~\bibnamefont
  {Yamamoto}}, \bibinfo {author} {\bibfnamefont {M.}~\bibnamefont {Koashi}}, \
  and\ \bibinfo {author} {\bibfnamefont {N.}~\bibnamefont {Imoto}},\ }\href
  {\doibase 10.1103/PhysRevA.64.012304} {\bibfield  {journal} {\bibinfo
  {journal} {Phys. Rev. A}\ }\textbf {\bibinfo {volume} {64}},\ \bibinfo
  {pages} {012304} (\bibinfo {year} {2001})}\BibitemShut {NoStop}%
\bibitem [{\citenamefont {Campbell}\ and\ \citenamefont
  {Benjamin}(2008)}]{PhysRevLett.101.130502}%
  \BibitemOpen
  \bibfield  {author} {\bibinfo {author} {\bibfnamefont {E.~T.}\ \bibnamefont
  {Campbell}}\ and\ \bibinfo {author} {\bibfnamefont {S.~C.}\ \bibnamefont
  {Benjamin}},\ }\href {\doibase 10.1103/PhysRevLett.101.130502} {\bibfield
  {journal} {\bibinfo  {journal} {Phys. Rev. Lett.}\ }\textbf {\bibinfo
  {volume} {101}},\ \bibinfo {pages} {130502} (\bibinfo {year}
  {2008})}\BibitemShut {NoStop}%
\end{thebibliography}%

\end{document}